\documentclass[superscriptaddress,amsmath,amssymb,aps,pre]{revtex4-2}

\usepackage{graphicx}
\usepackage{dcolumn}
\usepackage{bm}
\usepackage{xcolor}

\begin{document}

\preprint{APS/123-QED}

\title{Life in Complex Fluids: Swimming in Polymers}
\thanks{A footnote to the article title}%

\author{Paulo E. Arratia}%
 \email{parratia@seas.upenn.edu}
\affiliation{Department of Mechanical Engineering and Applied Mechanics, University of Pennsylvania, Philadelphia, PA 19104, USA}

\date{\today}

\begin{abstract}
Many microorganisms live and evolve in complex fluids. Examples include mammalian spermatozoa in cervical mucus, worms (e.g., \textit{C. elegans}) in wet soil, and bacteria (e.g., \textit{H. pylori}) in our stomach lining. Due to the presence of (bio)polymers and/or solids, such fluids often display nonlinear response to (shear) stresses including viscoelasticity and shear-rate dependent viscosity. The successful interaction between these microorganisms and their fluid environment is critical to the function of many biological processes including human reproduction, ecosystem dynamics, and the spread of disease \& infection. This interaction is often nonlinear and can lead to many unexpected behavior. Here, I will discuss developments in characterizing, modeling, and understanding the swimming behavior of model microorganism in viscoelastic and shear-thinning fluids. Three main microorganisms will be explored: (i) the nematode \textit{C. elegans}, an undulatory swimmer; (ii) the green algae \textit{C. reinhardtii}, a puller swimmer; and (iii) the bacterium \textit{E. coli}, a pusher swimmer. Investigation with artificial particles/swimmers will also be discussed; such studies are helpful in decoupling the biology from hydrodynamic effects. We will explore the interactions between these swimmers' gaits, geometry, and actuation and fluid rheological behavior using mostly experiments, and discuss these results relative to numerical and analytical predictions. 
\end{abstract}

\maketitle


\section{Introduction}

This manuscript is based on an invited talk at the 2021 American Physical Society-Division of Fluid Dynamics (APS-DFD) meeting. It was a return to an in-person setting, mostly, and a reminder that science is very much a human endeavor. Before we begin I would like to make it clear that this manuscript is not intended as a comprehensive review. For that, I guide the reader to many excellent treatments on the fundamentals of swimming at low Reynolds numbers (Re) \cite{lauga_Powers_2009, Goldstein_ARFM_Chlamy, Lauga_ARFM2016, Stocker_ARFM2012} and on motility of living organisms and propulsion of active particles in complex fluids \cite{Saverio_Underhill_ARFM, Arezoo_JNNFM, Arratia_ActiveColloids, RevModPhys.88.045006}. Rather, this article offers an experimentalist view on the current state of low-Re swimming in non-Newtonian fluids; technical details will only be briefly described and arguments will appear oversimplified, often relying on published literature. I hope such strategy does not jeopardize the reader's interest in the field; my goal is to provide a quick starting guide for those interested in joining our community. 

\begin{figure*}[h!t]
 \centering
 \includegraphics[height=4.75 cm] {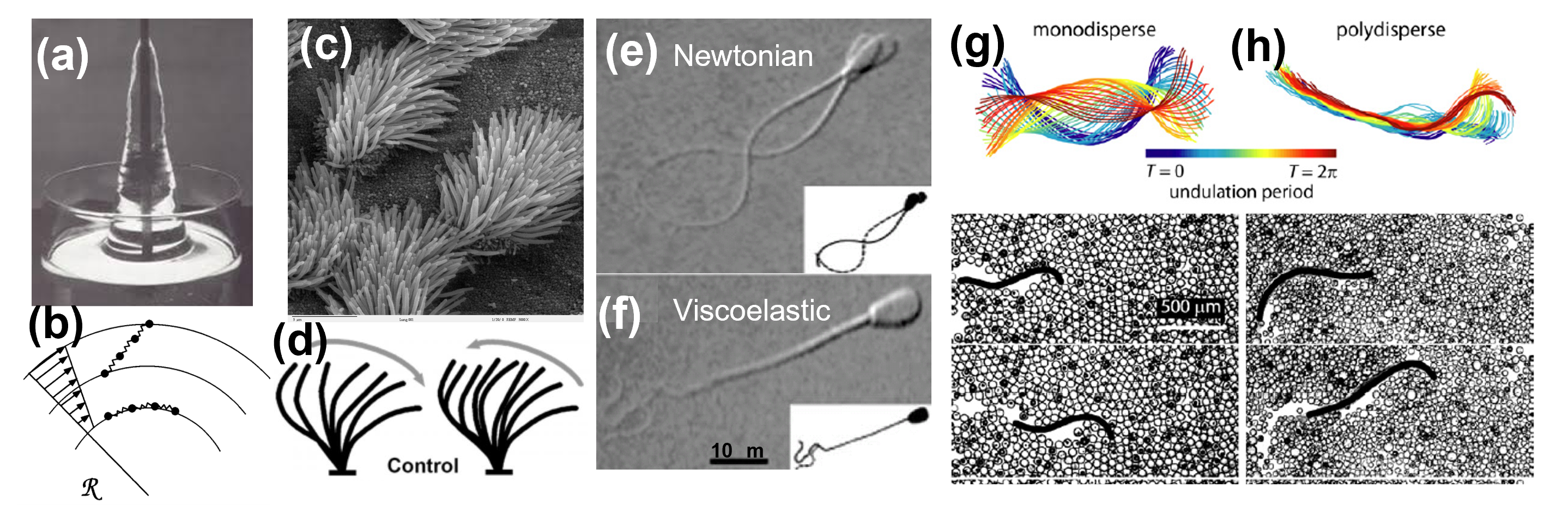}
 \caption{(a) Rod-climbing effect with a viscoelastic fluids (VE) \cite{bird_dynamics_1987}. (b) Schematic of polymer stretching in flows with curved streamlines \cite{pakdel_elastic_1996}. (c) Electron microscopy image of lung cilia (from Wikipedia). (d) Schematic of cilia normal beating cycle showing a power and recovery strokes \cite{Rossman_1984}. (e) Images of mammalian sperm cell moving in Newtonian. Inset show flagellum's nearly sinusoidal waveform. (f) Sperm swimming in VE fluids. (inset) Flagellum shows a hyper-extended waveform due to the presence of fluid elasticity  \cite{suarez_sperm_2005}. (g,h) The worm nematode \textit{C. elegan} moving in wet granular suspensions of (g) monodispersed and (h) polidispersed particles \cite{juarez_motility_2010}. Similar to sperm cells, interaction between nematode and suspension microstruture significantly affects kinematics.}
 \label{Fig:intro}
\end{figure*}

Complex fluids are widely found in nature and biology; examples include wet sand, mud, milk, cervical mucus, saliva, and blood. While homogeneous at the macroscale, these fluids often possess structure at an intermediate scale typically a few sizes of its constituents) Importantly, their macroscopic flow behavior (i.e., rheology) is a strong and nonlinear function of their microstructure \cite{larson_rheology_1999,Bonn_Yield_ARFM2017,guazzelli2011physical, graham2018microhydrodynamics, Ewoldt_ARFM2022, mckinley_filament-stretching_2002, galloway2022relationships, Morozov2015, SANCHEZ_Morozov_JNNFM_2022}. Many fascinating macroscopic responses of fluids containing polymer molecules, surfactants, colloids, liquid crystals, etc., have been reported in the literature over the years \cite{larson_instabilities_1992, shaqfeh_purely_1996,chen_shear_2000, arratia_elastic_2006, Brust_PlasmaPRL_2013, stone_dynamics_2006, van_hecke_topical_2010,morris2020shear}. In the particular case of polymeric fluids, the presence of (flexible) polymer molecules in the fluid and interaction of the molecules with the flow are responsible for nonlinear flow behavior such as hydrodynamic instabilities, drag reduction, and even turbulence \cite{giesekus_zur_1966, denn_fifty_2004, denn_issues_1990, bird_dynamics_1987, joseph_fluid_1990, shaqfeh_purely_1996, groisman_elastic_2000, arratia_elastic_2006, poole_purely_2007, groisman_efficient_2001, mckinley_wake_1993, datta_elastic_2022}. The exact mechanisms responsible for such phenomena are still being elucidated and is a topic of much current research \cite{datta_elastic_2022, SANCHEZ_Morozov_JNNFM_2022}. But we do know that mechanical stresses in these polymeric fluids are history dependent and depend on a characteristic time $\lambda$. In dilute solutions, this time scale has been found to be proportional to the relaxation time of a single polymer molecule \cite{schroeder_observation_2003, smith_response_1998, smith_single-polymer_1999}; in semi-dilute solutions, $\lambda$ depends also on molecular interactions \cite{larson_rheology_1999}. These (elastic) stresses grow nonlinearly with strain rate and can dramatically change the flow behavior. An example is the "rod-climbing" effect, in which a viscoelastic (VE) fluid (e.g., cake batter, bread dough, yogourt) creeps up a rod being rotated in the fluid \cite{bird_dynamics_1987} (Fig.~\ref{Fig:intro}a). This phenomenon was first described in the 1940's \cite{garner1946rheological} and involves a VE fluid being stirred by a rotating rod as shown in Fig.~\ref{Fig:intro}(a). The combination of high-velocity gradients and curved streamlines can stretch the (bio)polymer molecules, which leads to a normal stress difference $N1=\tau_{\theta\theta} - \tau_{rr}$, where $r$, $\theta$, and $z$ are cylindrical coordinates. This normal stress difference (or hoop stress since the rod is curved) produces a volume force, $N1/r$, that acts inwards against the outwards radial pressure gradient pushing the fluid up the rod (Fig.~\ref{Fig:intro}b). The development of such viscoelastic "hoop stresses", as polymer molecules are driven out of their equilibrium conformation by the imposed flow, induce radial secondary flows that is responsible for many destabilizing flow phenomena observed in VE flows \cite{ bird_dynamics_1987, joseph_fluid_1990, shaqfeh_purely_1996, groisman_elastic_2000, datta_elastic_2022, SANCHEZ_Morozov_JNNFM_2022}. As we will see here, these additional (elastic) stresses and time scales can significantly affect the swimming behavior of microorganisms.

Due to their small length scales, microorganisms such as bacteria, sperm cells, and various kinds of protozoa move/swim at low Reynolds ($Re$) (Fig.~\ref{Fig:intro}c-h). In such regime, fluid linear viscous forces dominate over nonlinear inertial ones \cite{brennen_fluid-mechanics_1977, childress_mechanics_1981, vogel_life_1994, lauga_hydrodynamics_2009}, and locomotion results from non-reciprocal deformations in order to break time-reversal symmetry; this is the so-called "scallop theorem" \cite{purcell_life_1977}. To survive, microorganisms must then seek locomotion strategies that break time-symmetry. Much work has been devoted in understanding such strategies in experiments, theory, and numerical simulations \cite{childress_mechanics_1981, vogel_life_1994, fauci_biofluidmechanics_2006, lauga_hydrodynamics_2009, Lauga_ARFM2016, Stocker_ARFM2012, Goldman_ARFM2015, Arratia_ActiveColloids}. Despite much progress, our understanding of swimming at low $Re$ numbers is mostly derived from investigations in Newtonian fluids. But there are many microorganisms (e.g., sperm cells, bacteria) that evolve in liquids that contain (bio)polymers, surfactants, and/or solids (e.g., mud, mucus, gels) \cite{fauci_biofluidmechanics_2006, Elfring2015, Sznitman2015, lauga_propulsion_2007} and exhibit non-Newtonian behavior such as shear-thinning viscosity and viscoelasticity. The question is: how these non-linear flow behaviors affect the swimming behavior of microorganisms?

Let us consider the cilia beating in the human lungs (Fig.~\ref{Fig:intro}c-d), which is lined with respiratory mucus. This complex biological fluid (mucus) has double duty: it protects against foreign particulates and pathogens while allowing the transport of gases and nutrients \cite{Mucus_Rheo_2009}. Not surprising, the rheological behavior of mucus is quite complex; it possesses large elastic and viscous modulus, which are both shear-rate dependent \cite{cone2005mucosal}. We can estimate the elastic effects on ciliary motion by computing the Elasticity number, $El=\lambda \mu/\rho L^2$. Here, $\mu$ are $\rho$ the fluid viscosity and density respectively, and $\lambda$ is the fluid relaxation time, and $L$ is a characteristic length scale associated with the microorganism. The quantity $El$ is often thought as the ratio of two time-scales: the time for elastic stresses to relax, $\lambda$, relative to the viscous time scale, $\rho L^2/\mu$. When $El \gg 1$, fluid elasticity dominates the dynamics. It is important to note that the $El$ is independent of flow kinematics or speed (i.e., $U$), and it is only a function of fluid properties and system geometry. For a typical cilia beating frequency ($\sim$ 60 hz), one can approximate mucus viscosity to be $\mu \sim \mathcal{O} (1)$ Pa$\cdot$s with a relaxation time $\lambda \sim \mathcal{O} (10)$s \cite{Mucus_Rheo_2009}. Taking the system length scale to be the flagellum length $L \sim \mathcal {O} (10^{-5})$m and $\rho \sim \mathcal{O}(10^3)$kg/m$^3$, we arrive at $El \sim \mathcal{O} 10^{10}$(!). This is an exceedingly high value demonstrates that lung ciliary motion occurs in a environment dominated by mucus elasticity. Note that $El$ scales inversely with the square of the organisms' length scale $L$ (usually in the $\mu$m scale), which means that elastic stresses are likely to be accentuated for micron-sized organisms. A prime example is the swimming of mammalian sperm cell (Fig.~\ref{Fig:intro}e,f), which switch from a nearly sinusoidal waveform in Newtonian liquids to a hyperextended waveform in elastic media \cite{suarez_sperm_2005,fauci_biofluidmechanics_2006}. 

The examples above illustrate the complex behavior once microorganisms encounter fluids with nonlinear rheological properties. The coupling between microorganisms' kinematics and fluid microstructure and the ensuing flow fields can give rise to unexpected results, some of which may seem counter-intuitive relative to Newtonian expectations. This coupling is often nonlinear and is a two way street: microorganism swimming motion affect the fluid response, and in turn the fluid affect the organisms's kinematics. Figure~\ref{Fig:intro} (g,h) shows how the waveform of small worm nematodes is affected by simply modifying the distribution of particle sizes (from monodisperse to polydisperse) in a granular suspension. Thus, from an experimental standpoint, it is important to work with model systems both in terms of choice of fluid and swimming microorganism; that is, fluid rheological properties (i.e., shear-thinning, viscoelasticity, yield-stress, etc) should be carefully characterized and organisms' velocity fields (in those fluids) should be measured and/or computed. There is a vast literature and well-established procedures for the former \cite{bird_dynamics_1987,Macosko_Rheology_1994,larson_rheology_1999, Ewoldt2015}, and the community is making significant strides in the latter. Some of those efforts will be discussed here.     

This article is organized as follows: Section II presents a brief background on the fundamental of swimming at low-Re in Newtonian and non-Newtonian fluids; Section II focuses on propulsion of artificial particles in complex fluids; Section III will discuss mainly experimental works on the swimming behavior of living microorganisms in viscoelastic fluids; Section IV provides a summary and outlook. 

\section{Brief Background}
We begin by briefly discussing the hydrodynamics of swimming at low $Re$; a more thorough review of the subject can be found elsewhere \cite{brennen_fluid_1977,Stocker_ARFM2012, lauga_hydrodynamics_2009}. Under steady, low $Re$ (no inertia) flow conditions, the equation of motion reduces to:
\begin{equation}
\label{Stokes}
    \nabla p = \nabla \cdot\bm{\tau},
\end{equation}
where $p$ is pressure and $\bm{\tau}$ is the deviatoric stress tensor. Equation~\ref{Stokes} is known as the Stokes' equation, named after the mathematician Sir George Stokes \cite{brennen_fluid-mechanics_1977}. For Newtonian fluids, the stress $\bm{\tau}$ is linearly proportional to the strain rate $\dot{\bm{\gamma}}$ such that $\bm{\tau} = \mu \dot{\bm{\gamma}}=\mu \left(\nabla \mathbf{u}+(\nabla \mathbf{u})^{T}\right)$, where the constant of proportionality is the dynamic viscosity $\mu$. Equation~(\ref{Stokes}) can then be expressed as:
\begin{equation}
    \label{Stokes_N}
    \nabla p = \mu \nabla^{2}\mathbf{u}.
\end{equation} 

Note that Equation~(\ref{Stokes_N}) is linear in both velocity $\mathbf{u}$ and pressure $p$. Equation~(\ref{Stokes_N}) is also instantaneous; it has no dependence on time other than via boundary conditions. The lack of time dependence means that the flow is reversible. An external force, $\mathbf{F}(t)$, will lead to a flow that upon reversal of the force, $\mathbf{F}(-t)$ and its history, brings the flow back to its original state. This kinematic reversibility forms the hydrodynamic basis of the scallop theorem put forth by Purcell in 1977 \cite{purcell_life_1977}. These hydrodynamic properties illustrate that swimming at low Re can seem at first as a highly confined phenomenon, yet microorganisms have found a variety of ways to overcome the constraints of the scallop theorem.

But what if a microorganism is instead swimming in a complex fluid, as in the case of sperm cells in cervical mucus \cite{suarez_hyperactivated_2003} (Fig.~\ref{Fig:intro}f). Such fluids display a plethora of nonlinear rheological behavior including yield stress, thixotropy, shear-thinning viscosity behavior, and viscoelasticity. To describe such flow behavior one needs to develop constitutive models that can accurately capture the nonlinear relationship between (deviatoric) stress ($\bm{\tau}$) and strain-rate ($\dot{\bm{\gamma}}$). That, of course, is easier said than done and much effort has been devoted to the development of constitutive models for complex fluids \cite{bird1976useful, bird1995constitutive, Randy_ARFM, alves2021numerical}. Here we will briefly discuss two such instances: shear-thinning viscosity and viscoelasticity. 

\subsubsection{Shear-Thinning Fluids} 

\begin{figure*}[h!t]
 \centering
 \includegraphics[height=6cm] {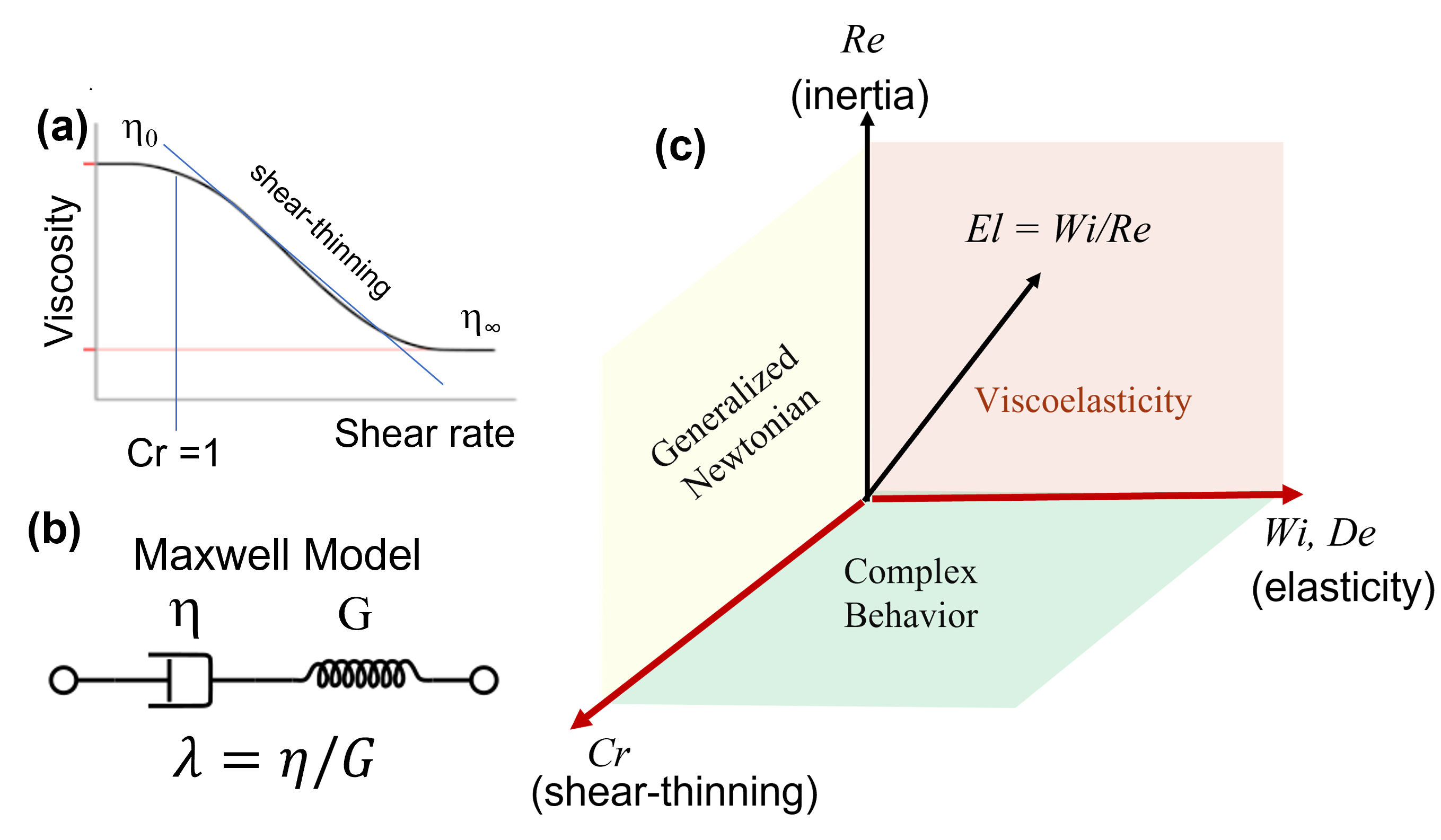}
 \caption{Schematic of fluid rheological models and working phase space. (a) Sketch of shear-thinning viscosity behavior captured by Carreau-Yassuda model. The Carreau number, $Cr$ describes the transition from Newtonian to power-law behavior. (b) Schematic of the Maxwell model, a linear constitutive equation for viscoelasticity. Its described by an element containing a viscous damper and an elastic spring connected in series. (c) Non-dimensional phase space. This manuscript will focus on low Reynolds number behavior in which inertia is virtually negligible. Two main fluid behavior will be explored: viscoelasticity and shear-thinning viscosity behavior. The Weissenberg ($Wi$) \& Deborah ($De$) numbers characterize elasticity, while the Carreau ($Cr$) number characterizes shear-thinning.}
 \label{Fig:PhaseSpace}
\end{figure*}

Many biological fluids exhibit shear-rate dependent viscosity (e.g., shear-thinning and shear-thickening), that is $\bm\tau = \eta(\dot{\gamma}) \bm{\dot{\gamma}}$, where $\dot{\bm{\gamma}}=\left(\nabla \mathbf{u}+(\nabla \mathbf{u})^T\right)$ is the strain-rate tensor and $\eta(\dot{\gamma})$ is a non-Newtonian viscosity. Note that $\dot{\gamma}$ is the magnitude of the strain-rate tensor, $\dot{\gamma} =  \sqrt{1/2 (\bm{\dot{\gamma}} \colon \bm{\dot{\gamma})}}$. Shear-thinning fluids have a viscosity that decreases as shear rate increases (e.g., paints, ketchup), while shear-thickening fluids posses a viscosity that increases as shear rate increases (e.g., suspensions of corn starch). This non-Newtonian viscosity is often described by an empirical power law model of the type $\eta(\dot{\gamma})=k|\dot{\gamma}|^{n-1}$, where $k$ is a viscosity factor and $n$ is a power law index. If $n>1$, the fluid is shear thickening whereas if $n<1$ the fluid is shear-thinning; for $n=1$, the model reduces to Newtonian behavior. This viscosity model, however, is unbounded in the limit of low ($\dot{\gamma} \rightarrow 0$) and high ($\dot{\gamma} \rightarrow \infty$) shear rates, producing nonphysical viscosity values in those limits. A more realistic (empirical) model for shear-thinning fluids is the Carreau-Yassuda viscosity model \cite{bird_dynamics_1987} usually given as:
\begin{equation}
\label{Carreau}
\eta=\eta_{\infty}+(\eta_{0}-\eta_{\infty})[1+\lambda_{c}|\dot{\gamma}|^{a}]^{(n-1)/a},
\end{equation}
where $\eta_0$ is the zero-shear viscosity, $\eta_{\infty}$ is the infinite-shear viscosity, and $n$ is the usual power-law index. The quantity $\lambda_{c}$ is a time-scale associated with the shear-rate (in the unit of inverse time) at which the fluid viscosity departs from Newtonian behavior. When the exponent $a=2$, then the equation above is known as the Carreau model. We can define a Carreau number, $Cr=\lambda_{c} \dot{\gamma}$, which is used to characterize the transition from Newtonian-like behavior (zero-shear rate region) and power-law region; $Cr=1$ marks the departure from low shear-rate Newtonian viscosity (see Fig.~\ref{Fig:PhaseSpace}a for a schematic). This model offers advantages over the power law model discussed above. The most significant perhaps is that the Carreau-Yassuda model is able to capture the frequently observed viscosity transition from a low-shear-rate Newtonian plateau to the power-law region as the shear-rate $\dot{\gamma}$ is gradually increased. 

\subsubsection{Viscoelastic Fluids}
Fluid elastic stresses accumulate, grow nonlinearly with strain rate, and are expected to significantly affect the swimming behavior of microorganisms. Accurately describing such stresses (state and history) is, however, quite the challenge \cite{morozov_introductory_2007, datta_elastic_2022}. A simple, linear constitutive model is the Maxwell model, which is represented by a viscous damper and a elastic spring connected in series \cite{Fung_1994,larson_rheology_1999} (Fig.~\ref{Fig:PhaseSpace}b). The Maxwell model is usually expressed as:
\begin{equation}
\label{Maxwell}
    \bm\tau +\lambda \frac{d\bm\tau}{dt} = \eta \bm{\dot{\gamma}},
\end{equation}
where $\lambda=\eta/G$ is the fluid relaxation time and $G$ is the spring elastic modulus, This model introduces a time-dependent stress that is proportional to "fluid elastic memory" $\lambda$, and it reverts to Newton's law of viscosity for $\lambda=0$. The Maxwell model predicts that the stress relaxes exponentially in time, which is relatively accurate for many dilute polymeric solutions. However, Eq.~(\ref{Maxwell}) predicts that stress will increase linearly with time under constant stress, a trend not observed in rheological measurements. Importantly, the Maxwell model is only valid for small deformations. For large deformations, one can generalize the Maxwell model by incorporating frame-invariance, which leads to the upper-convected Maxwell model in the following tensorial form:
\begin{equation}
\label{upper-conv Maxwell}
    \bm{\tau} +\lambda \overset{\nabla}{\bm{\tau}} = \eta \dot{\bm{\gamma}}.
\end{equation}
Here, $\overset{\nabla}{\bm{\tau}}$ denotes the upper-convected derivative of $\bm{\tau}$, defined as:
\begin{equation}
    \overset{\nabla}{\bm{\tau}}= \frac{\partial\bm\tau}{\partial t}+\mathbf{u}\cdot\nabla\bm{\tau}-(\nabla \mathbf{u})^T\bm{\tau}-\bm{\tau}(\nabla \mathbf{u}).
\end{equation}
While a significant improvement over its linear counterpart, the the upper-convected Maxwell (UCM) model does not consider the contribution from the solution's solvent viscosity ($\eta_s$) to the total stress; hence it fails to predict the ``retardation effect'' of elasticity when a step change in stress is applied. The Oldroyd-B model addresses this and other issues \cite{datta_elastic_2022, SANCHEZ_Morozov_JNNFM_2022}, as is usually written as:
\begin{equation}
\label{Oldroyd-B}
\bm{\tau} +\lambda \overset{\nabla}{\bm{\tau}} = \eta( \dot{\bm{\gamma}}+\lambda_r\overset{\nabla}{\dot{\bm{\gamma}}}),
\end{equation}
where $\lambda_r=\lambda\eta_s/(\eta_p+\eta_s)$ is the fluid retardation time, and $\eta_s$ and $\eta_p$ are the viscosities of the solvent and the polymer, respectively. A parameter $\beta=\eta_s/(\eta_p+\eta_s)$ is usually defined, and one recovers the UCM model in the limit of zero solvent viscosity, $\eta_s=0$. While the Oldroyd-B model is quite useful, it also has its limitations: it cannot capture rate-dependent viscosity and normal stress behaviors, and its stresses become unbounded in extensional flows beyond a critical (extensional) rate.  These issues arise mostly due to the infinite extensibility of the model polymer chains -- the Finite-Extensibility Nonlinear Elastic (FENE) type-models can address (some of) these issues. Nevertheless, the Oldroyd-B model is often employed to simulate viscoelastic shear flows where stretching is relatively moderate \cite{smith_single-polymer_1999} and is known to capture many nontrivial viscoelastic phenomena such as the development of hoop-stresses and hydrodynamic instabilities~\cite{larson_purely_1990, shaqfeh_purely_1996, bird_dynamics_1987, datta_elastic_2022, SANCHEZ_Morozov_JNNFM_2022}. It is important to note the nonlinear relationship between stress, $\bm{\tau}$, and flow field, $\mathbf{u}$, and its consequence for swimming studies. It indicates that swimmers with different motility kinematics such as body undulations and rotation of helical flagella are expected to produce different responses from the (viscoelastic) fluid.

One can define two main dimensionless parameters to describe the effects of elasticity. The first is the Deborah number ($De$), defined as the ratio of fluid relaxation time to the flow time scale such that $De=\lambda/T$. Here $T$ is characteristic time-scale associated with the flow deformation process; fluid-like behavior is obtained in the limit of $De=0$. In swimming studies, this flow time-scale is often substituted by the microorganisms' beating frequency $f$ such that $De=f\lambda$. The second is the Weissenberg number ($Wi$) which quantifies the degree of nonlinearity associated with (fluid) normal stresses $N_1=2\lambda\eta\dot{\gamma}^2$ (from UCM) relative to shear stresses $\tau = 2\eta\dot{\gamma}$; hence $Wi=\lambda\dot{\gamma}$. (For more information, please see \cite{poole2012deborah}.) Nonlinear elastic stresses are expected to become important in the flow for $De,Wi>1$. 

A working phase space can now be defined using the dimensionless numbers describe here (Fig.~\ref{Fig:PhaseSpace}c). The different axis quantify the effects of inertia ($Re$), elasticity, and shear-thinning viscosity ($Cr$). These forces/effects often appear together along the planes in the phase space, and such situations have yet to be studied in detail. Here, however, we will focus on cases in which fluid inertia in negligible, that is, $Re\ll 1$. We will straddle the $Wi,De$ and $Cr$ axis, focusing on situations in which one of these effects is dominant.

\section{Reciprocal swimmers: Can fluid rheology enable propulsion?}

\begin{figure*}[h!t]
 \centering
 \includegraphics[height=6cm] {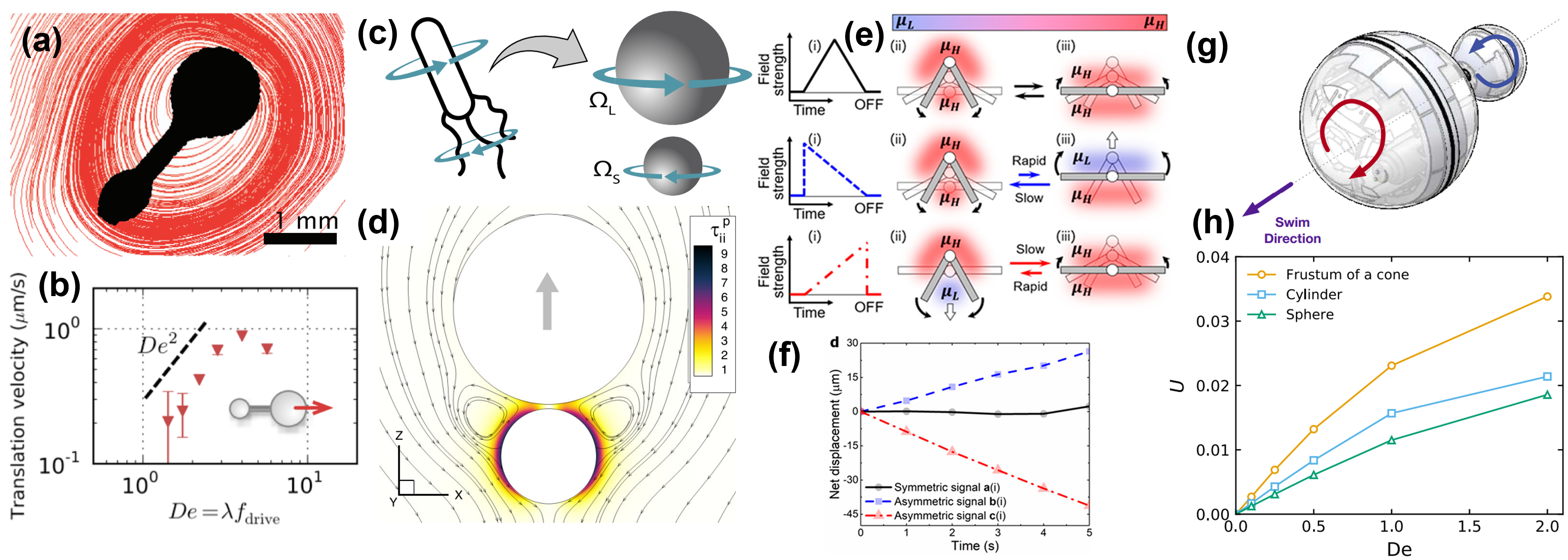}
 \caption{Breaking the scallop theorem in non-Newtonian fluids. (a,b) Asymmetric dimer ($L \approx 3$mm) being reciprocally actuated in a viscoelastic (VE) fluid at $Re \ll 1$. The dimer produces enough curvature in the streamline to generate normal stresses in fluid. Dimer propulsion speed increases (quadratically) with $De$ \cite{keim_fluid_2012} (c,d) Counter-rotating spheres in VE fluids can achieve propulsion due to hoop stresses surrounding the faster-spinning smaller sphere \cite{Binagia_PRFluids_2021}. (e,f) Self-assembled, magnetic colloidal scallops moving in shear-thinning fluids; propulsion directionality can be controlled by tuning the actuation and/or colloidal structure size~\cite{Velev_Langmuir2020}. (g,h) A autonomous robotic swimmer based on (c,d) \cite{Shaqfef_ArXiv_2022}.}
 \label{Fig:particles}
\end{figure*}

The discussion above makes it clear that fluid rheology can significantly affect the swimming behavior of living organisms. But can fluid rheology enable propulsion at low $Re$? That would mean breaking the scallop theorem \cite{purcell_life_1977}, for which the main assumptions are no inertia and Newtonian behavior (see Eq.~\ref{Stokes_N}). If we relax these assumptions, then it may be possible to break kinematic reversibility and achieve net motion even for reciprocal swimming strokes \cite{lauga_life_2011}. Consider Purcell's scallop now immersed in a shear-thinning fluid. If the scallop opens and closes its mouth at different rates during one stroke, then it may produce different non-Newtonian shear viscosities ($\eta$) (in space and time) along the stroke's path if the condition of $Cr>1$ is met. That is, the viscosity field around the scallop would be non-uniform with the lowest viscosity values near the boundary and largest values away from boundary at a particular instant in the stroke. This would mean that the scallop would experience different viscous stresses during one reciprocal stroke. This viscous stress ``imbalance" may be enough to lead to net motion. That is, (shear) rates matter when it comes to propulsion in shear-rate-dependent viscosity fluids. The possibility that fluid rheological properties could enable propulsion has been explored for a handful of special cases: a flapping surface extending from a plane~\cite{normand_flapping_2008, pak_pumping_2010}; a sphere which generates small-amplitude sinusoidal motion of fluid along its surface~\cite{lauga_hydrodynamics_2009}; a ``wriggling'' cylinder with reciprocal forward and backward strokes at different rates~\cite{fu_swimming_2009}; oscillating \cite{Pak_PoF2012, Elfring_2018} and counter-rotating spheres \cite{Binagia_PRFluids_2021} (Fig.~\ref{Fig:particles}~c,d). Analysis of the flow fields generated by these "swimmers" moving in Oldroyd-B and FENE-P model fluids suggests that elastic effects can generate forces that scale quadratically with the amplitude of the motion \cite{normand_flapping_2008, pak_pumping_2010}. This demonstrates that fluid elastic stresses can be exploited to enable propulsive forces, circumventing the scallop theorem.

Nearly a decade ago, fluid-assisted propulsion for a reciprocal swimmer was experimentally demonstrated in viscoelastic fluids~\cite{keim_fluid_2012,Gagnon_2014}. In those studies, a single rigid object, in this case an asymmetric dumbbell particle or dimer, is externally actuated in a reciprocal manner in viscous fluids. In the experiments, the dimer such as the one shown in Fig.~\ref{Fig:particles}(a) is repeatedly reoriented by a magnetic field. The effects of inertia are absent due to the high fluid viscosity ($\sim 10$~Pa$\cdot$s); the $Re \approx 10^{-4}$, a value comparable to that of a swimming microorganism. By applying only magnetic torques, the apparatus reciprocally actuates just one degree of freedom in the system, the dimer's orientation $\hat a$. No net motion is observed for the Newtonian case since $\hat a(t)$ is cyclic; this is as expected. Yet when a small amount of polymer \cite{keim_fluid_2012} or surfactants \cite{Gagnon_2014} are added to the Newtonian solvent (corn syrup), the same cyclic stroke results in net propulsion in a direction set by the dimer's shape and boundary conditions.

Figure~\ref{Fig:particles}b shows dimer speed as a function of $De$ for dilute polymeric solutions \cite{keim_fluid_2012}. The first observation is that the dimmer speed increases monotonically as $De$ increases; that is, the more elastic the fluid becomes, the faster the dimer propels itself. The dimer speed seems to obey a $De^{2}$ scaling or $U \sim (f\lambda)^2$. It is worth noting that at low frequencies, assuming Oldroyd-B fluid model, $G' \sim f^2$ where $G'$ is the fluid elastic modulus. Thus, the observed propulsion seems to be a purely elastic effect, likely generated by the interaction of polymer molecules with the flow curved streamlines (Fig.~\ref{Fig:particles}a). Similar to the rod-climbing effect (see Fig.~\ref{Fig:intro}a,b), the combination of polymer stretching with flow velocity gradients and curved streamlines generated by the actuated dimer lead to a volume force (or ``hoop stress") $N_1/r$. Because of the dimer is asymmetric, so are these hoop stresses, and that imbalance leads to the dimer's net motion. These stresses are history-dependent and do not entirely cancel out over one forcing period, but instead have a small rectified component that accumulates particularly as $De$ increases.  

Propulsion may also be enabled by other fluid rheological properties, such as shear-rate dependent viscosity. Indeed, Qiu \emph{et al.} \cite{qiu2014swimming} have observed net propulsion in reciprocally-swimming micro-scallops immersed in shear-thinning and shear-thickening fluids in both experiments and simulations. Recently, it has been shown that one can manipulate not only the speed but also the direction of propulsion of re-configurable magnetic "colloidal scallops" by carefully controlling the actuation rates in shear-thinning fluids (Fig.~\ref{Fig:particles}e,f) \cite{Velev_Langmuir2020}. The direction of propulsion changes with both the size and structure of these colloidal assemblies because of the different viscous stresses that they produce and experience. This viscous imbalance is thought to be responsible for particle propulsion for $Cr>1$. 

In the experimental investigations described above, the "swimmers" or particles are externally actuated. That is, they are force but not torque free. Very recently, however, a robotic autonomous dimer particle has been developed specifically for propulsion in viscoelastic media by local hoop stresses \cite{Binagia_PRFluids_2021} (Fig.~\ref{Fig:particles}g,h). Remarkably, the robotic system passively adapts to propel itself forward at different speeds, depending on the properties of the surrounding fluid. As a result, this prototype can serve as a local rheometer for complex fluids environments allowing the estimation of quantities such as first and second normal stress differences \cite{Shaqfef_ArXiv_2022}. The passive sensing capability of this robotic swimmer can lead to many application in biology and human health. 

In summary, there is growing evidence that fluid nonlinear rheological properties can be exploited to break the scallop theorem and obtain propulsion for artificial swimmers. Such swimmers can move through complex fluids with only reciprocal actuation, a simple body shape, and/or no moving parts -- a less complicated design than other propulsive strategies. Experiments with artificial particles are also helpful in decoupling the biology from hydrodynamic effects \cite{zenit_PoF2013, liu_force-free_2011,zenit_ST_2017}, which permits a more direct comparison with analytical works. It is important to note, however, that just because kinematic reversibility is broken, it does not mean that one has achieved efficient propulsion; it only means that propulsion is possible. For example, the dimmers described in \cite{keim_fluid_2012, Gagnon_2014} have propulsive efficiencies ($\mathcal{O}$(1\%)) similar to those of non-reciprocal swimmers in Newtonian fluids, including magnetic torque-driven helical micro-robots ($\approx 1\%$ \cite{Peyer_microrobots}) and self-propelling force-free bacteria ($\approx 2\%$ \cite{Wu_PNAS_2006}). That is, there is still much room for improvement. Further understanding of factors controlling this efficiency could greatly simplify fabrication of micro-swimmers in many complex, artificial environments or for biological settings where non-linear rheology is ubiquitous.

\section{Swimming of Microorganisms in Complex Fluids}
We now turn our attention to studies with living microorganisms. Emerging studies - some of which are discussed in \cite{Arezoo_JNNFM,Saverio_Underhill_ARFM}- are revealing the effects of fluid rheology on the swimming behavior of microorganisms. The goal is to understand the nonlinear coupling between the microorganism's swimming kinematics and fluid rheological properties. To do so, at least experimentally, it is advisable to work with model systems, both in the choice of microorganism and working fluid. Model organisms are non-human species which are extensively studied to understand particular biological phenomena and/or function. Examples include the zebra fish (\textit{Danio rerio}), the bacterium \textit{Escherichia coli}, fruit fly (\textit{Drosophila melanogaster}), and the nematode \textit{Caenorhabditis elegans}, among many others. The idea is that discoveries made in model organisms will provide insight into the workings of other non-model organisms, including humans. In the case of swimming studies, the vast wealth of genetic information available for these systems allows precise control over their motility strategies. From a hydrodynamics's standpoint, one would also wish to work with organisms for which their kinematics and the velocity fields have been characterized, at least in the base case (Newtonian fluids). 

\begin{figure*}[h!t]
 \centering
 \includegraphics[height=7.5cm] {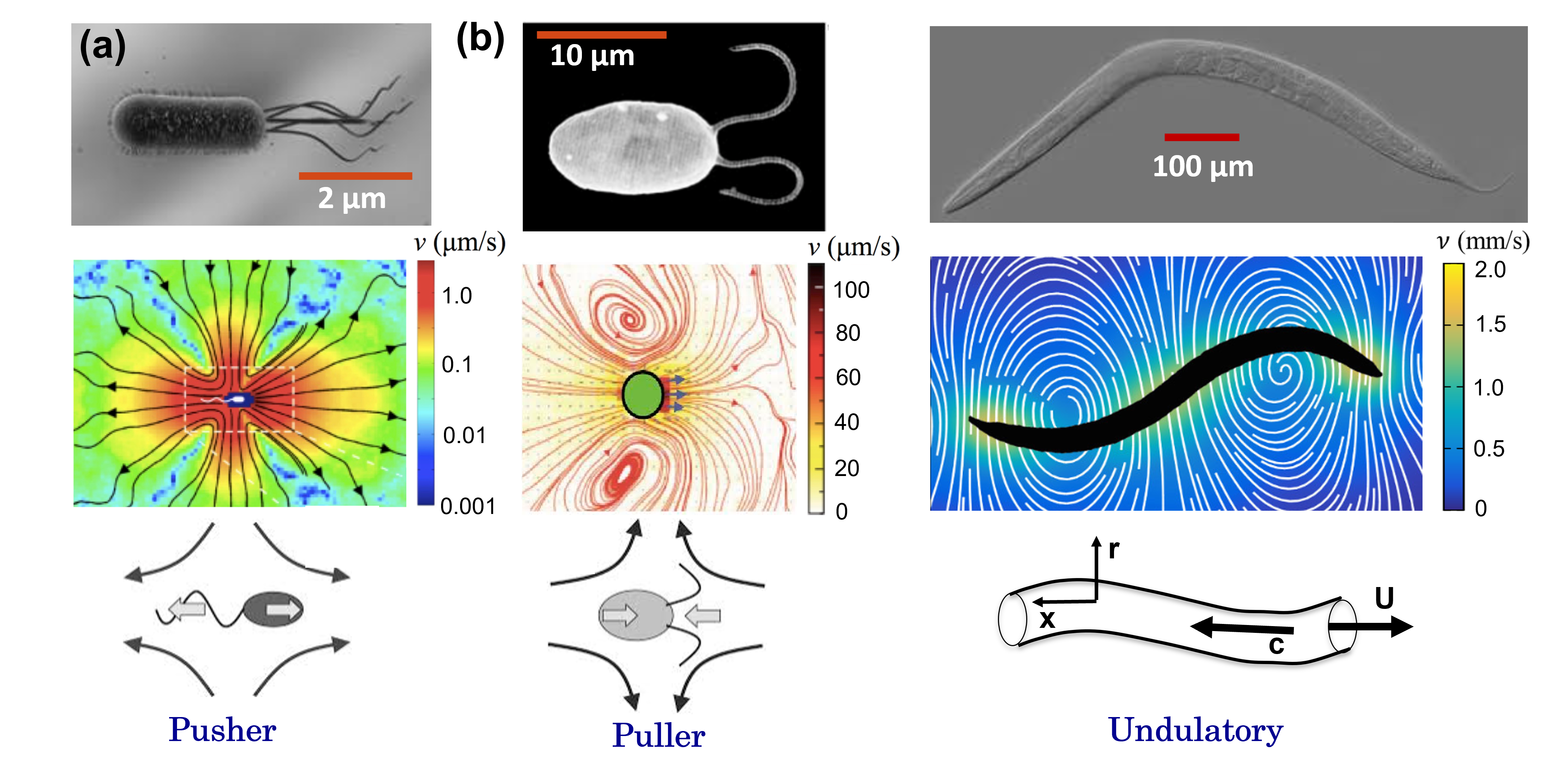}
 \caption{Model organisms for swimming studies. (a) The bacterium \textit{E. coli}, (b) the green algae \textit{C. reinhardtii}, (c) The nematode \textit{C. elegans}. Below each organism is their corresponding, time-averaged experimentally measured velocity field. The flow data indicates that (a) behaves as a pusher \cite{Drescher_Ecoli_2011}, (b) as a puller \cite{drescher_direct_2010}, and (c) as an undulatory swimmer \cite{gagnon_lauga_PRF}} 
 \label{Fig:Swimmers}
\end{figure*}

Three main model organism for swimming studies will be discussed here: the nematode \textit{C. elegans}, the green algae \textit{C. reinhardtii}, and the bacterium \textit{E. coli}. The swimming kinematics and the resulting velocity fields for all these organisms are well established in Newtonian fluids \cite{sznitman_propulsive_2010, Drescher_Ecoli_2011, drescher_direct_2010}. Based on these data, one can consider \textit{C. elegans} to be a model undulatory swimmer that resembles Taylor's waving cylinder, while \textit{C. reinhardtii} and \textit{E. coli} are considered to be "puller" and "pusher" swimmers, respectively (Fig.~\ref{Fig:Swimmers}). We will discuss these classifications in more detail shortly. 

Equally important is to develop model fluids with known rheological properties. Ideally, working fluids should emphasize a single rheological behavior such as elasticity or shear-thinning viscosity behavior. Such model fluids have been extensively used in the field of rheology and non-Newtonian fluid mechanics~\cite{larson_rheology_1999, shaqfeh_purely_1996, datta_elastic_2022}. An example is the well-known "Boger" fluid, developed by David Boger in the 1970's \cite{boger_highly_1977, BogerFLuids_ARFM_2009}. This fluid is highly elastic but maintains a nearly-constant shear viscosity. These rheological features are approximated by the Oldroyd-B constitutive model (partially), and thus Boger fluids have been widely used in the study of the effects of viscoelastic on fluid flows. Unfortunately, some of the polymer and specially solvents involved in the formulation of typical Boger fluids are toxic to many microorganisms, limiting its application for swimming and biological studies. Alternatives do exist, but adequately characterizing the rheological properties of the working fluids is critical. 

\subsection{Undulatory Swimming: From Taylor's waving sheet to \textit{C. elegans}}

\subsubsection{Purely Viscous Fluids} 
\begin{figure*}[h!t]
 \centering
 \includegraphics[height=3 cm] {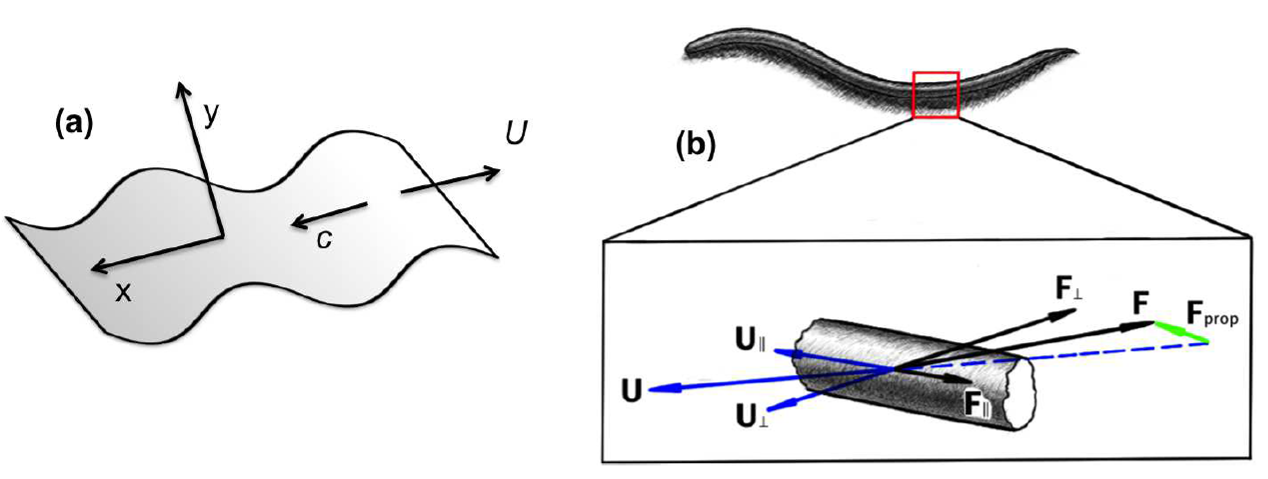}
 \caption{Schematic of two-dimensional waving sheet in a viscous fluid illustrating the traveling wave of velocity $c$ progressing in the x-direction and the forward swimming speed ($U$) in opposite direction.
(b) Application of Resistive Force Theory (RFT) on \textit{C. elegan} body illustrating the normal and tangential components of the velocity $U$ and force $F$, and the resulting net propulsive force.}
\label{Fig:RTF}
\end{figure*}

Nearly seventy years ago, G. I. Taylor \cite{Taylor51,Taylor52} beautifully demonstrated that a slender body such as a (non-extensible) waving sheet (Fig. \ref{Fig:RTF}a) could swim in an incompressible, Newtonian fluid by generating traveling waves in the absence of inertia. The sheet oscillates in time according to $y(x,t)=b\sin(kx-\omega t)$, where $b$ is the traveling wave amplitude, $\omega$ is the frequency, and the wavelength $\Lambda$ is $2\pi/k$; the traveling wave speed is $c=\omega/k$. For vanishing $Re$, the sheet oscillations induce a speed $U_N=0.5\omega b^2 k+O(kb)^4$ \cite{Taylor51}. That is, if the fluid is at rest relative to the sheet, then the sheet is propelled in the direction opposite to that of the propagation of the distorting wave. Note that fluid properties, such as viscosity, do not enter in Taylor's speed equation for the waving sheet. Taylor later considered the case of waving cylindrical tails, in which waves of lateral displacement move down a filament \cite{Taylor52}. While the analysis is limited to small amplitudes and fixed kinematics, it provided one of the first predictions regarding the propulsion in viscous environments. It should be noted that around the same time J. Lighthill showed that a deformable body could move in a viscous fluid with a speed proportional to the square of the deformation amplitude \cite{Lighthill_1952_Squirm}. Soon after (1953), G. J. Hancock \cite{hancock_self-propulsion_1953} (a student of Lighthill) built on Taylor's results but took a different approach: he distributed Stokes' singularities, \textit{Stokelets} and dipoles, along a waving filament's center-line, which was the starting point for the well-known slender body theory (SBT)~\cite{Pedley_Lighthill_ARFM}.

Many important investigations followed these pioneering works. Of particular importance is the introduction of resistive force theory (RFT) developed by Gray and Hancock \cite{Gray55}. RFT assumes that the hydrodynamics forces are proportional to the local body velocity such that the force exerted by a body or flagellar segment is given by $\mathbf{F}= C_N \mathbf{U_N} + C_T \mathbf{U_T}$. Here, $C$ corresponds to the local drag coefficient per unit length that depends on geometry and fluid viscosity, and $N$ and $T$ are the normal and tangential components, respectively (see Fig.~\ref{Fig:RTF}b). It is the anisotropy between the normal and tangential drag coefficients, with $C_N>C_T$, that lies at the origin of drag-based thrust; for infinitesimally thin filament, Gray and Hancock found $C_n/C_t =2$. While RFT is only an approximate solution (each element is independent of the other), it has been widely applied with good success in biological systems \cite{johnson_flagellar_1979,ishijima_RTF_2011, lauga_hydrodynamics_2009}, and even in granular systems \cite{Goldman_Science2009, Goldman_PoF2014, Goldman_ARFM2015}. 

Later, Lighthill \cite{Lighthill76} re-introduced and extended the the viscous slender body theory (SBT) presented in Hanckock's 1953 manuscript \cite{hancock_self-propulsion_1953} to improve RFT by pointing out importance of long-range hydrodynamics interactions and incorporation slender body approximations. Such improvements led to $C_N/C_T=1.5$ for the case of an undulating filament moving in an unbounded fluid medium. Experiments with \textit{C. elegans} found very similar values with $C_N/C_T \approx 1.4$~\cite{sznitman_propulsive_2010}. When incorporating wall-effects into the analysis, a significantly larger value of the drag coefficient ratio ($C_N/C_T = 4.1$) is obtained \cite{katz_1974}; that is, the propulsive speed is faster near walls. SBT formulation has become almost standard for the analysis of undulatory swimmers at low $Re$, and many excellent analytical and numerical works have emerged since then \cite{childress_mechanics_1981, vogel_life_1994, brokaw_simulating_2001, fauci_biofluidmechanics_2006, lauga_propulsion_2007, korta_mechanosensation_2007, avron_optimal_2004, shapere_self-propulsion_1987, johnson_flagellar_1979, dresdner_propulsion_1980, wiggins_flexive_1998, gauger_numerical_2006, lauga_Powers_2009}. A major challenge, however, is to extend this framework to fluids that displays both solid and fluid-like behavior, such as viscoelastic fluids. 
 
\subsubsection{Undulatory Swimming in Complex Fluids}
One of the first attempts to incorporate the effects of fluid elasticity on undulatory swimming used a series of expansions similar to Taylor’s analysis and a second-order fluid constitutive model\cite{Chaudhury_JFM79}. The analysis show that fluid elasticity could either increase or decrease self-propulsion depending on the value of $Re$. It is important to note that the second-order fluid model is a (second-order) asymptotic approximation about the rest state of a given viscoelastic fluid and is only valid for slow and slowly varying velocity fields. Thus its applicability to Taylor's waving sheet problem is probably inadequate. Later, inspired by observations of spermatozoa swimming in mucus \cite{katz_movement_1978,katz_movement_1981}, the effects of elasticity on beating flagella were considered using the (linear) Maxwell model \cite{fulford_swimming_1998} (see also Eq.~\ref{Maxwell}). It was shown that self-propulsion was not affected by fluid elasticity even at large Deborah numbers ($De$), but the total work decreased with increasing $De$. These results should be interpreted carefully since the Maxwell model is not valid for large deformations. 

About 15 years ago, Lauga \cite{lauga_propulsion_2007} showed that, for a 2D waving sheet (Fig. 2a), elastic stresses could significantly alter the organism speed and the work required to achieve net motion. Using nonlinear constitutive models (e.g., Oldroyd-B, FENE-P), the author showed that the organism speed ($U$) is given by the equation

\begin{equation}
	\frac{U}{U_N}=\frac{1+\beta De^2}{1+De^2}.
   	\label{Lauga}
\end{equation} 

The Newtonian velocity is defined as $U_N$, which is Taylor’s original result. In Eq. \ref{Lauga}, $\beta=\eta_s/(\eta_p+\eta_s)$ is the ratio of the solvent viscosity to the solution (total) viscosity, as defined in Eq.~\ref{Oldroyd-B}. Hence, for a given swimming gait $U<U_N$; that is, elastic stresses reduces the swimmer overall speed relative to the Newtonian base case. Fu, Wolgemuth, and Powers found similar expression for the case of 2D waving cylinder or filament \cite{fu_theory_2007} and extended to 3D finite-size bodies \cite{fu_swimming_2009}. Unlike Taylor's result, Eq. \ref{Lauga} depends on fluid material properties; similar to Taylor's result, Eq. \ref{Lauga} is for a given (fixed) kinematics. In reality an organism could compensate the reduction in velocity by increasing its beating frequency and/or decrease in wavelength. Nevertheless, Eq. \ref{Lauga} represents an important step forward since it provides a quantitative measure of the effects of fluid elasticity on the swimming speed of microorganisms. It spurred much activity in the field, some of which we will discuss here. 

Numerical simulations have also been used to address the role of fluid elasticity on the swimming behavior of microorganisms. In particular, Teran, Fauci, and Shelley~\cite{teran_viscoelastic_2010} considered two-dimensional swimming "free" sheets (i.e., with free head and tail) of \textit{finite} length $L$ in an Oldroyd-B fluid. The simulations show that, for accentuated tail motions, the sheet swims faster at $De \approx 1$ than in a Newtonian fluid, that is, "swimmer" stroke frequency matches the fluid relaxation time. As elasticity is increased, the filament swimming speed decreases as predicted by Eq.~(\ref{Lauga}). Further developments show that swimmer speed could increase or decrease in viscoelastic fluids depending on swimmer gait/kinematics \cite{Riley_2014,ELFRING_JNNFM_2016} as well as the filament material properties (e.g., stiffness) \cite{Thomases_PRL_2014}. But do experiments corroborate these findings?

\subsubsection{Experiments with \textit{C. elegans}}
The nematode \textit{Caenorhabditis} (\textit{C}.) \textit{elegans} is a multi-cellular, free-living roundworm found in soil environments. The nematode posses a quasi-cylindrical body shape of length $L \approx 1$ mm and radius $r \approx 80 \mu$m. Much is known about the nematode's genetics and physiology; its genome has been completely sequenced~\cite{brenner_genetics_1974} and its cell lineage has been established~\cite{byerly_life_1976}. These nematodes are equipped with 95 muscle cells that are highly similar in both anatomy and molecular makeup to vertebrate skeletal muscle~\cite{White_1986}. Their neuromuscular system controls their body undulations which allows \textit{C. elegans} to swim, dig, and crawl through diverse environments. The wealth of biological knowledge accumulated to date makes \textit{C. elegans} ideal candidates for investigations that combine aspects of biology, biomechanics, and the fluid mechanics of propulsion. This slender nematode can serve as experimental analog of Taylor's waving cylinder problem \cite{Taylor52, fu_theory_2007}.

\begin{figure*}[h!t]
 \centering
 \includegraphics[height=7.5 cm] {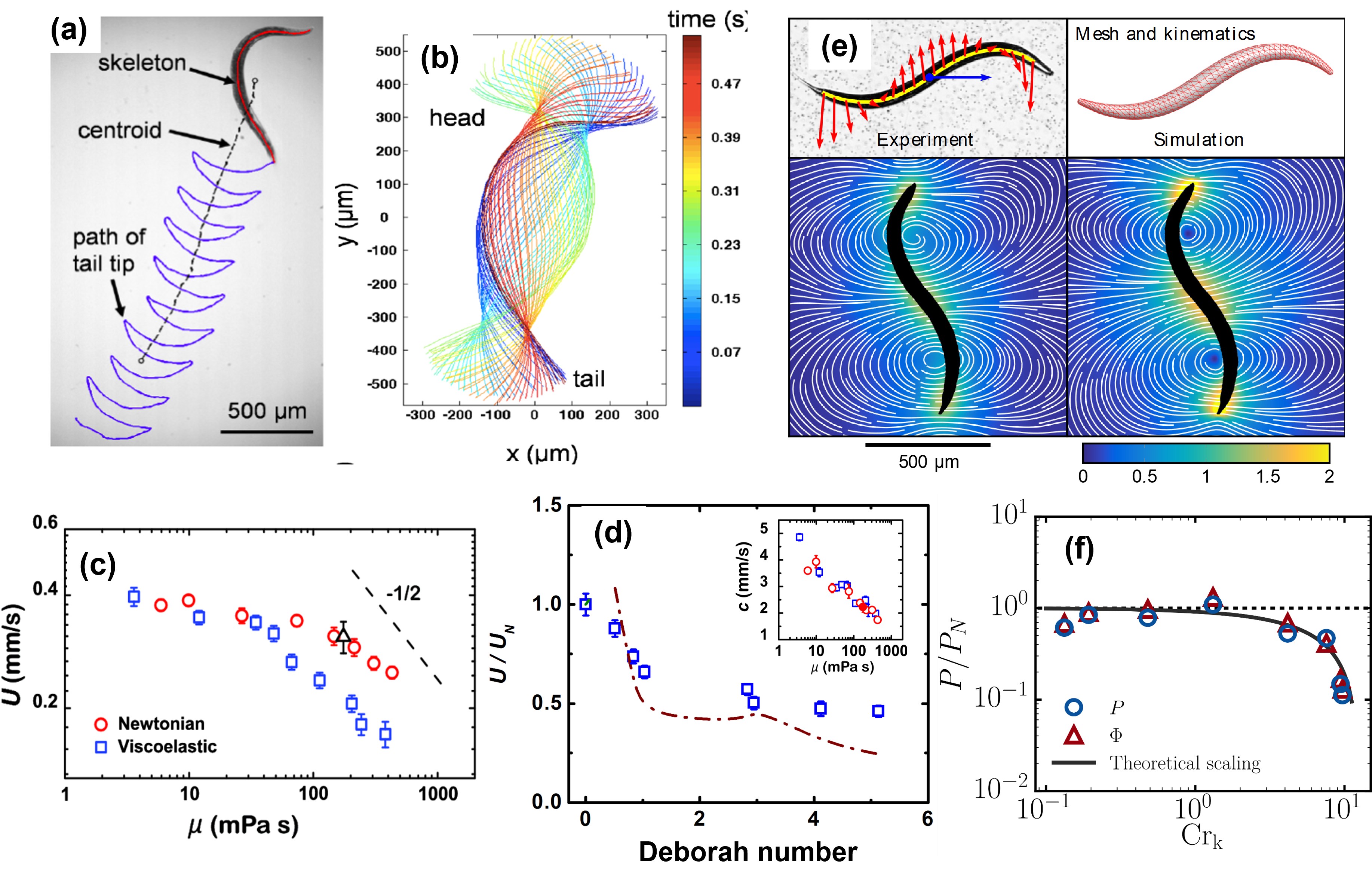}
 \caption{The nematode \textit{C. elegans} swimming in complex fluids. (a) bright-field image of nematode moving in Newtonian fluids over several beating cycles. The nematode's centroid path shows periodic body oscillations. (b) Nematode's centerline postures over one beating cycle measured using image analysis and color coded by time. (c) Nematode swimming speed, $U$ as a function of fluid viscosity for both Newtonian and VE fluid cases. (d) Normalize swimming speed as a function of $De$. Inset shows that nematode wavespeed is affected by fluid viscosity but not by polymers. (e) Experimental and numerical velocity fields \cite{gagnon_lauga_PRF}. (f) Normalized power expenditure by swimming nematode in shear thinning fluids. Nematode spends less power in shear thinning fluids for $Cr>1$.}
 \label{Fig:Elegans}
\end{figure*}

Figure~\ref{Fig:Elegans}a shows an image of the nematode moving in a Newtonian buffer solution together with the path of its body centroid over multiple beating cycles. Due to its size ($L\sim$1 mm), the swimming nematode can be imaged using standard bright-field microscopy, an experimental advantage over microscopic systems. The nematode swimming speed ($U$) is calculated by differentiating the nematode centroid position with respect to time. Figure~\ref{Fig:Elegans}b shows the nematode's body (centerline) postures as a function of time, obtained via image analsys. Note that the amplitude is larger at the nematode's head than at its tail, indicative of the traveling waves moving along the nematode's body \cite{sznitman_propulsive_2010, shen_undulatory_2011}. 

An important consideration in swimming experiments with live organisms is the fluid medium. Fluids must be constructed or developed such that they possess the desirable rheological property (elasticity, shear-thinning, etc) but without being toxic to the organism. In the case of \textit{C. elegans}, Newtonian fluids of different shear viscosities, $\mu$, are prepared by mixing two low molecular weight oils (Halocarbon oil, Sigma-Aldrich), while viscoelastic fluids are prepared by adding small amounts of carboxymethyl cellulose (CMC, $M_w=7~\times 10^{5}$) into de-ionized water. By varying the polymer concentration in solution, one can tune the level of elasticity in the fluid and obtain fluid relaxation times $\lambda$ ranging from 0.4 s for the most dilute concentration (1500 ppm) to about 5.6 s for the most concentrated solution (8000 ppm). This strategy provides a reasonable dynamic range in fluid elasticity (about an order of magnitude). These CMC solutions are not purely elastic, they display shear-thinning viscosity behavior too. In order to compensate for the effects of shear-rate dependent viscosity, an aqueous solution of the stiff polymer Xanthan Gum (XG) that is shear-thinning but possesses negligible elasticity is used in experiments; more details can be found at \cite{shen_undulatory_2011}.  

\textit{Propulsion Speed: Newtonian vs Viscoelastic}: With the methods in place, it is now possible to address the question of whether fluid elasticity hinders or enhances the propulsion speed of live organisms. The nematode's swimming speed as a function of fluid viscosity for both Newtonian and viscoelastic (CMC) solutions is shown in Fig.~\ref{Fig:Elegans}(c). For relatively low viscosity values, the swimming speed is independent of $\mu$ and the values of $U$ are nearly identical for both cases. For $\mu > 30$ mPa$\cdot$s, however, the swimming speed ($U$) decreases with increasing $\mu$ even for Newtonian fluids; recall that Taylor's result for the waving sheet is independent of fluid properties \cite{Taylor51, Taylor52}. This decrease in $U$ is most likely due to the nematode's finite power. The speed data shows that $U$ decays slower than $\mu^{-1/2}$, suggesting that the nematode does not swim with constant power. Importantly, the values of $U$ for viscoelastic fluids are found to be 35 \% lower than the Newtonian fluid of same shear viscosity (Fig.~\ref{Fig:Elegans}c). Thus, it appears that fluid elasticity hinders propulsion speed of an undulatory swimmer in agreement with analytical results \cite{lauga_propulsion_2007, fu_theory_2007}. 

The effects of fluid elasticity on the nematode's swimming behavior are best illustrated by plotting the normalized swimming speed $U/U_{N}$ as a function of the Deborah number ($De=\lambda f$), as in Eq.~\ref{Lauga}. Here, $U_{N}$ is the Newtonian speed. Figure~\ref{Fig:Elegans}(d) shows that the normalized swimming speed decreases monotonically with $De$, and reaches an asymptotic value of 0.4 as $De$ is further increased. That is, it appears that elastic stresses introduces resistance to propulsion, therefore decreasing the nematode's swimming speed; more details in \cite{shen_undulatory_2011}. The experimental data seems to agree relatively well with analytical predictions \cite{lauga_propulsion_2007, fu_theory_2007} and Eq.~\ref{Lauga}. Of course, such agreement is not necessarily expected because there are significant differences between the experiments and the calculations. For example, the analysis are two-dimensional (2D) while the nematode is allowed to swim in 3D although only planar swimming was considered in the experiments. Most importantly, while the calculations imposes a particular prescribed kinematics or waveform, the nematode is free to choose its own. In fact, we find that the nematode's wavespeed, $c$, decreases as a function of fluid viscosity for both Newtonian and VE cases, as shown in Fig.~\ref{Fig:Elegans}(d, inset). Nevertheless, the agreement is rather remarkable and may point to generic features in this problem.

So what could explain the decrease in swimming speed for nematodes moving in viscoelastic fluids? We may find clues in the velocity fields produced by the swimming nematodes. Figure~\ref{Fig:Elegans}(e) shows experimental and numerical velocity fields for swimming \textit{C. elegans} in Newtonian fluids \cite{gagnon_lauga_PRF}; the numerical velocity field was obtained using boundary element methods (BEM) along with time-resolved nematode's body postures obtained in experiments. The agreement between the numerical and experimental velocity fields is quite remarkable, and it allow us to inspect the base flow. A common feature of the velocity fields are regions of fluid recirculation that are aligned along the nematode’s body. These recirculation regions persist throughout the bending cycle, but their exact location varies. The flow structures presented here and elsewhere \cite{sznitman_propulsive_2010} show that the nematode's velocity field are complex and does not strictly fall into the pusher-puller category. The velocity fields also show curved streamlines and high velocity gradients, which can locally stretch polymer molecules and lead to the production of extra elastic stresses. These extra stresses can lead to additional resistance to propulsion and hinder swimming speed. In fact, it was originally thought that these elastic stresses were produced in extensional regions of the flow, which in turn dramatically increase the local extensional viscosity of the medium. Numerical simulations, however, showed that not to be the case; rather they found large elastic stresses produced near the head of swimmer where the amplitude is higher \cite{Thomases_PRL_2014}.  

While much effort has been devoted to understand the effects of fluid elasticity on swimming behavior of undulatory swimmers, shear-thinning effects has received much less attention. That is an oversight since shear-thinning behavior is very common in polymeric solutions. Using Taylor's waving sheet along with a Carreau fluid model, V\'elez-Cordero and Lauga~\cite{Velez-Cordero_2013} showed that the "swimmer" is more efficient in the shear-thinning fluid even though its speed remains the same as in the Newtonian case. A numerical simulation by Montenegro-Johnson, Smith, and Loghin~\cite{montenegro_2013} showed that for large amplitude waves the swimming speed increases in shear-thinning fluids. These recent studies have shown that even relatively simple non-Newtonian fluid behavior can have a significant impact on the swimming kinematics of microorganisms. In experiments, shear-thinning viscosity seems to have little to no effect on the swimming speed of \textit{C. elegans} (as in \cite{Velez-Cordero_2013}), but it modifies the velocity fields produced by the swimming nematode \cite{gagnon_shearThining_2014}. Velocimetry data show significant enhancement in local vorticity and circulation. Figure~\ref{Fig:Elegans}(f) shows that the work or cost of swimming required for nematodes to move in shear-thinning fluids is less than that of a purely viscous fluids for $Cr >1$ \cite{gagnon_shearThining_2014, li_ardekani_2015}. So, it may be "easier" for \textit{C. elegans} to swim in shear-thinning fluids. 

In summary, experiments with the nematode \textit{C. elegans} shows that fluid elasticity hinders its swimming speed \cite{shen_undulatory_2011} while shear thinning viscosity had no effect on $U$ \cite{gagnon_shearThining_2014}. The data indicates that the more elastic the fluid is, the slower the nematode will swim (until an asymptote is reached). This trend is predicted by both numerical simulations \cite{teran_viscoelastic_2010, Thomases_PRL_2014} and theory \cite{lauga_propulsion_2007, fu_theory_2007}, but the agreement is only qualitative. There is still room for refining both experiments and analysis, particularly in resolving time-dependent, 3D flows. For example, it is still unclear how the nematode's body material properties (tissue viscosity, body elasticity and bending stiffness) couples with fluid rheology \cite{Thomases_PRL_2014, sznitman_effects_2010, lauga_eloy_2013} and the ensuing swimmer kinematics. Experiments with \textit{C. elegans} swiming in viscous fluids show that nematode's Young's modulus and tissue viscosity increase as fluid viscosity increases \cite{sznitman_effects_2010}, and simulations describe how soft swimmers "soft" filaments can swim faster than stiffer ones, a result corroborated (at least in part) by an analysis of the Taylor swimming sheet \cite{Riley_2014}. The governing dimensionless parameter is the Sperm number defined as $Sp=(\eta\omega/\kappa k^3)^{1/3}$, where $\eta$ is the fluid total viscosity, $\omega$ is the swimmer beating frequency, $k$ is wave number, and $\kappa$ is the swimmer bending stiffness. Fluid stresses are negligible for $Sp<< 1$ (stiff limit), but they become increasingly important as $Sp$ increases beyond unity. The challenge before us is to understand and describe how the (active) nematode's kinematics emerge from the interactions with its fluid environment. The idea would be to incorporate neuro-activity and -muscular models and data into swimming models to understand how observed \textit{C. elegans}'s motility behavior is related to sensory inputs \cite{Samuel_CPNeuro_2009,Gagnon_eNeuro2018}. 

\subsection{Pulling \& Pushing in Complex Fluids}
We now turn our attention to two archetypal modes of swimming, namely pusher and pullers (Fig. \ref{Fig:Swimmers}a,b). These types of swimmers are a mathematical construct (from Eq.~\ref{Stokes_N}) developed to describe the flow field generated by real microorganisms. As noted by Hankock \cite{hancock_self-propulsion_1953}, at Low $Re$ flow disturbances driven by the kinematic motion of a swimming microorganism depend linearly upon the stresses exerted by the moving body on the fluid; the velocity fields of such flow disturbances are described as linear superpositions of fundamental solutions of the Stokes' equation and decay with inverse powers of $r$ (or swimmer length scale). The first solution, referred to as a ``Stokeslet'', arises from the net force on the fluid and decays as $1/r$. The next solution, known as a ``stresslet'' flow, is induced by the first force moment exerted by the body on the fluid and decays more rapidly ($1/r^2$); higher-order solutions decay even more rapidly ($1/r^3$). The combination of these basic solutions can yield flows with complex and qualitatively different behaviors, exhibiting contrasting near- and far-field behaviors \cite{Drescher_Ecoli_2011, drescher_direct_2010, guasto_oscillatory_2010}.

Consider, for example, a neutrally buoyant force-free micro-swimmer propelling itself along its axial direction $\mathbf{e}$ (unit vector). The swimmer produces a force dipole $\mathbf{p}=\alpha \mathbf{e}$ in the fluid. Two different types of force dipoles can in general arise. Swimmers described by a negative force dipole ($\alpha<0$) are called “pullers” (Fig. \ref{Fig:Swimmers}b); they draw fluid in along the elongated direction and push fluid out from the sides. The actuation for pullers occurs near the particle head, and a prime example is the algae \textit{Chlamydomonas reinhardtii} (Fig. \ref{Fig:Swimmers}b). Swimmers described by positive force dipole ($\alpha>0$) are called “pushers” (Fig. \ref{Fig:Swimmers}a) in the sense that they repel fluid from the body along their axis and draw fluid in to the sides. A pusher swims by an actuating stress along the posterior of its body, and examples include the bacteria \textit{Escherichia coli} (Fig. 1b) and \textit{Bacillus subtilis}. Then, broadly speaking, the kinematics of microorganisms can be classified into two main types: pushers (\textit{E. coli}) and pullers (e.g. \textit{C. reinhardtii}). Note that other organisms such as the alga Volvox carteri may fall between this pusher/puller distinction. While this pusher-puller classification is limited and simplified, it provides a dichotomy for a reasonable framework.

\subsubsection{Pulling in Viscoelastic Fluids}
\begin{figure*}[h!t]
 \centering
 \includegraphics[height=7cm] {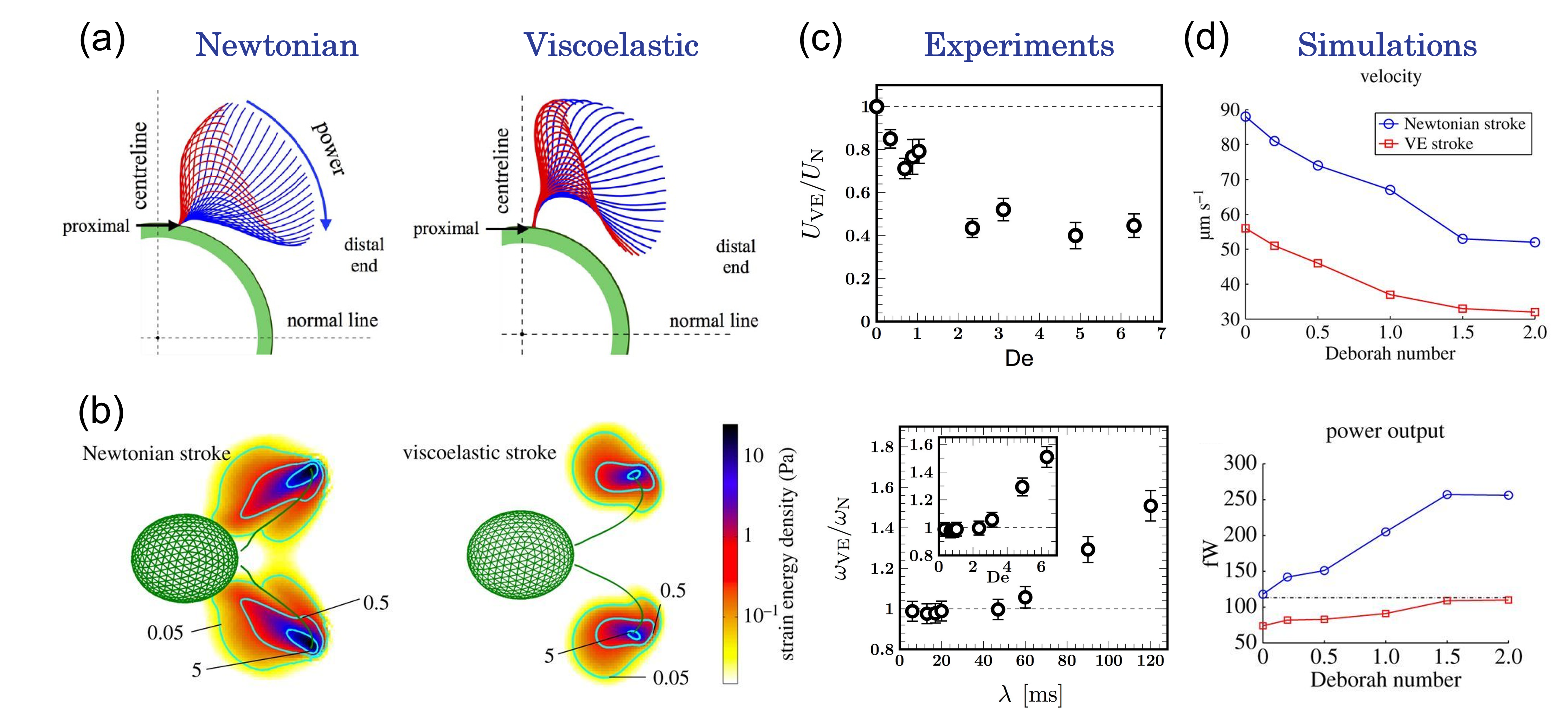}
 \caption{The puller swimmer \textit{C. reinhardtii} swimming in Newtonian and viscoelastic fluids. (a) Time-averaged flagellum strokes over one beating cycle. Fluid viscoelasticity dramatically constricts the the algal flagellar beating waveform, compared to the control case (b) Strain energy density (polymeric stresses) for \textit{C. reinhardtii} for both Newtonian and VE strokes at $De=2$. Note that Newtonian stroke produces significantly more elastic stresses than the VE stroke. (c) Fluid elasticity hinders cell swimming speed (top) but leads to a sharp increase in beating frequency (bottom). (d) Combination of experiments and simulations show that Newtonian stroke moves faster than a VE stroke in VE fluids (top) but spends more energy (bottom).}
 \label{Fig:Chlamy}
\end{figure*}

The alga \textit{Chlamydomonas reinhardtii} is a model system in biology and has been widely used in studies of motility \cite{Goldstein_ARFM_Chlamy}. The algae has ellipsoidal cell body that is roughly 10 $\mu$m in size and two anterior flagella each of length $L \approx 10 \mu$m, Structurally, the two flagella possess the same conserved ‘‘9 + 2’’ microtubule arrangement seen in other organisms axonemes including mammalian sperm cells. The algae executes a cyclical breaststroke-like patterns with asymmetric power and recovery stokes at frequencies $f \approx 30-60$ Hz to generate propulsion (Fig.~\ref{Fig:Chlamy}a). This swimming gait generates far field flows corresponding to an idealised \textit{puller} \cite{drescher_direct_2010, Goldstein_ARFM_Chlamy}.

Motile cilia and flagella are important sensors of their environment. The dynein-dependent sliding of microtubules and subsequent relaxation that governs the bending of cilia and flagella can be significantly affected by the characteristics of the external fluidic environment such as viscosity and elastic stresses. Increasing fluid viscosity, for example, can activate Ca2+ influx pathways that, in turn, increase cilia beating frequency \cite{ishijima_flexural_1994}. Thus, one expects extracellular conditions to affect motor function and flagellar waveforms. For example, sperm flagellum shows high amplitude waveforms in low viscosity fluids, while relatively rigid waveforms with large tail amplitudes are found in high viscosity fluids \cite{ishimoto:JTB:2018:sperm_sim_from_data}. Notably, experiments by Susan Suarez and colleagues have shown that fluid non-Newtonian rheology can significantly modify mammalian sperm flagellar kinematics \cite{suarez_hyperactivated_2003, fauci_biofluidmechanics_2006}, which translates into faster swimming speed and enhanced ability to penetrate the vestments encasing the egg \cite{suarez_sperm_2005}.

Recently, the effects of fluid elasticity on flagellar kinematics and cell motility has been systematically investigated in experiments \cite{Qin_SciRep_2015} and simulations \cite{qin_chuanbin_2017} using the green alga \textit{C. reinhardtti}. Experiments are performed in a thin fluid film is order avoid issues with solid boundaries. Two main fluids are used: (i) a Newtonian buffer solutions and (ii) viscoelastic (VE) polymeric solutions. The dilute polymeric solutions are aqueous solutions of high molecular weight polyacrylamide, a flexible polymer. By carefully varying polymer concentration, we can construct solutions with relaxation time, $\lambda$, ranging from 6 ms to 0.12 s. That translates into Deborah numbers ($De=f\lambda$) ranging from 0.3 to 6 if one takes $f=50$Hz; that is, over one order of magnitude in elastic effects. Newtonian fluids with viscosity values ranging from 1 cP to 10 cP (10x the viscosity of water) are produced to investigate purely viscous effects. Similar to the experiments with \textit{C. elegans}, we compare results from VE and Newtonian fluids at similar viscosity values to isolate elastic from viscous effects. More information about methods and fluid characterization can be found in \cite{Qin_SciRep_2015}.

Results show that \textit{C. reinhardtii} flagellar kinematics is significantly affected by both viscous and elastic stresses (Fig.~\ref{Fig:Chlamy}a); these modified kinematics in turn affect fluid flow and stress fields (Fig.~\ref{Fig:Chlamy}b). Both flagella' beating frequency and cell swimming speed decrease as fluid viscosity increases (not shown, see Fig. 2 in \cite{Qin_SciRep_2015}). That is, the cell swimming kinematics is significantly affected by even linear viscous stresses, similar to \cite{ishijima_flexural_1994}. Figure~\ref{Fig:Chlamy}a shows that the shape beating forms (over one cycle) for the Newtonian case differs from the VE case even at similar viscosity values ($\mu=6$ cp). The VE case shows flagellar movement severely restricted (less mobile) or bundled together near the cell body. Most of the bending or movement seems to occur away from the cell body with large localized bending at the distal tips. How does these kinematic changes translate into cell swimming speed?

Similar to \textit{C. elegans}, the cell normalized swimming speed ($U/U_N$) decreases as $De$ (or fluid elasticity) increases, as shown in the top panel of Fig.~\ref{Fig:Chlamy}(c). The decay of $U/U_N$ vs $De$ resembles the theoretical predictions and Eq.~\ref{Lauga} \cite{lauga_propulsion_2007,fu_theory_2007}. But the agreement may be coincidental since the cell beating frequency $f$ is found to increase, sharply, for $De \ge 2$, as shown in the bottom panel of Fig.~\ref{Fig:Chlamy}(c). This is curious since one would expect the swimming speed to increase as $\omega$ increases, illustrating the non-trivial response of cell flagellum to external fluid stresses. Unfortunately, experiments alone are insufficient to fully understand this non-trivial response and the governing mechanism that lead to a particular waveform. Is the emerging waveform a result of a passive response solely based on the material properties of the flagellum? Or is it an active response based on motor response to external load? Or a combination of both?

\subsubsection{Numerical Simulations with Pullers in VE fluids} 
Numerical simulations \textit{together with experimental data} can provide information that goes beyond what can be experimentally measured  or numerically calculated alone. In particular, it can provide data on local polymeric/elastic stresses and insights into the question of flagellum active vs passive response to flow stresses. To that end, Becca Thomases, Bob Guy, and colleagues recently developed the first 3D numerical model \cite{qin_chuanbin_2017} of a micro-organism swimming in a complex fluid with swimming kinematics derived solely from experimental data (Fig.~\ref{Fig:Chlamy}b). A numerical tool was designed to prescribe the exact kinematics (obtained from experiments) to separate the effects of gait and fluid rheology. Numerical simulations were validated by comparing the resulting swimming speed from the simulations to those from experiments. With the model and methods in place, it is possible for the first time to visualize the elastic stress accumulation in the fluid medium and to measure the energy expended by the \textit{C. reinhardtii}. 

To isolate the effect of fluid elasticity on swimming behavior and flagellar kinematics, experimental data on the gaits of \textit{C. reinhardtii} swimming in Newtonian and VE fluids are used as inputs to numerical simulations. Thus, simulations are able to independently change swimming kinematics (gait/stroke) and fluid rheology (viscosity/elasticity). Figure~\ref{Fig:Chlamy}(b) shows polymeric (elastic) stress fields (elastic strain energy density) for Newtonian and VE strokes beating in a VE fluid at $De=2$. Both strokes are obtained from experiments at similar viscosity values ($\approx 2.5$ cP); they are then placed/immersed in a numerical VE fluid simulated using the Oldroyd-B model (Eq. \ref{Oldroyd-B}. Results show that most of the polymeric stresses are produced along the flagellum and near the distal tips; polymeric stresses are relatively low around the alga's body. Surprisingly, the simulations show that the Newtonian stroke induces higher elastic stress than its VE counterpart, as shown in Fig. \ref{Fig:Chlamy}(b). These elevated stresses are responsible for the larger power needed by the swimmers using the Newtonian stroke to propel in VE fluids, even though they swim faster (Fig.~\ref{Fig:Chlamy}d). That is, the VE stroke is more energy efficient (but slower) suggesting that the swimmer may change its stroke (or gait) to the fluid properties based on energy availability. 

In summary, these results show that fluid material properties, in particular viscoelasticity, can significantly affect flagellar kinematics (stroke) and cell speed \cite{Qin_SciRep_2015, qin_chuanbin_2017}. The mechanism responsible for observed changes in kinematics are still unclear. Numerical simulations suggest, however, that such changes are an active response but we still do know to what extend or the precise mechanisms. On the other hand, these findings suggest that one may control the ciliary/flagellar beating and tune transport properties (e.g., cleareance of mucus) by manipulating fluid rheology. This opens up the possibility of using ciliary response to fluid properties to treat airway disease related to impaired cilia motility, such as primary cilia dyskinesia and cystic fibrosis, where “thickened” mucus due to large amount of DNA, actin, and bacterial biofilms leads to reduced mucociliary clearance and breathing difficulty.

\subsubsection{Running \& Tumbling in Polymeric Fluids}

As we have seen so far, the two-way coupling between swimmer kinematics and fluid rheological properties can give rise to many unexpected behaviors for microorganism swimming in complex fluids. We now explore the case of the bacterium \textit{E. coli}, an archetypical model organism for motility studies \cite{Lauga_ARFM2016}. \textit{E. coli} are rod-shaped cells (1 to 2 $\mu$m in size) with 3 to 4 helical flagella that rotate and bundle together as the they swim forward at speed of approximately 10 $\mu$m/s (in buffer solution). Notably, \textit{E. coli} moves using run-and-tumble dynamics that is diffusive at long times \cite{berg_bacterial_1975, Lauga_ARFM2016}; the "run state" is characterized by forward swimming while the "tumble state" is characterized by changes in cell direction due to motor reversal. Their velocity field is well-approximated by an idealised \textit{pusher} \cite{Drescher_Ecoli_2011}.

\begin{figure*}[h!t]https://www.overleaf.com/project/62c1ae992a6bff77973ddbff
 \centering
 \includegraphics[height=8.2cm] {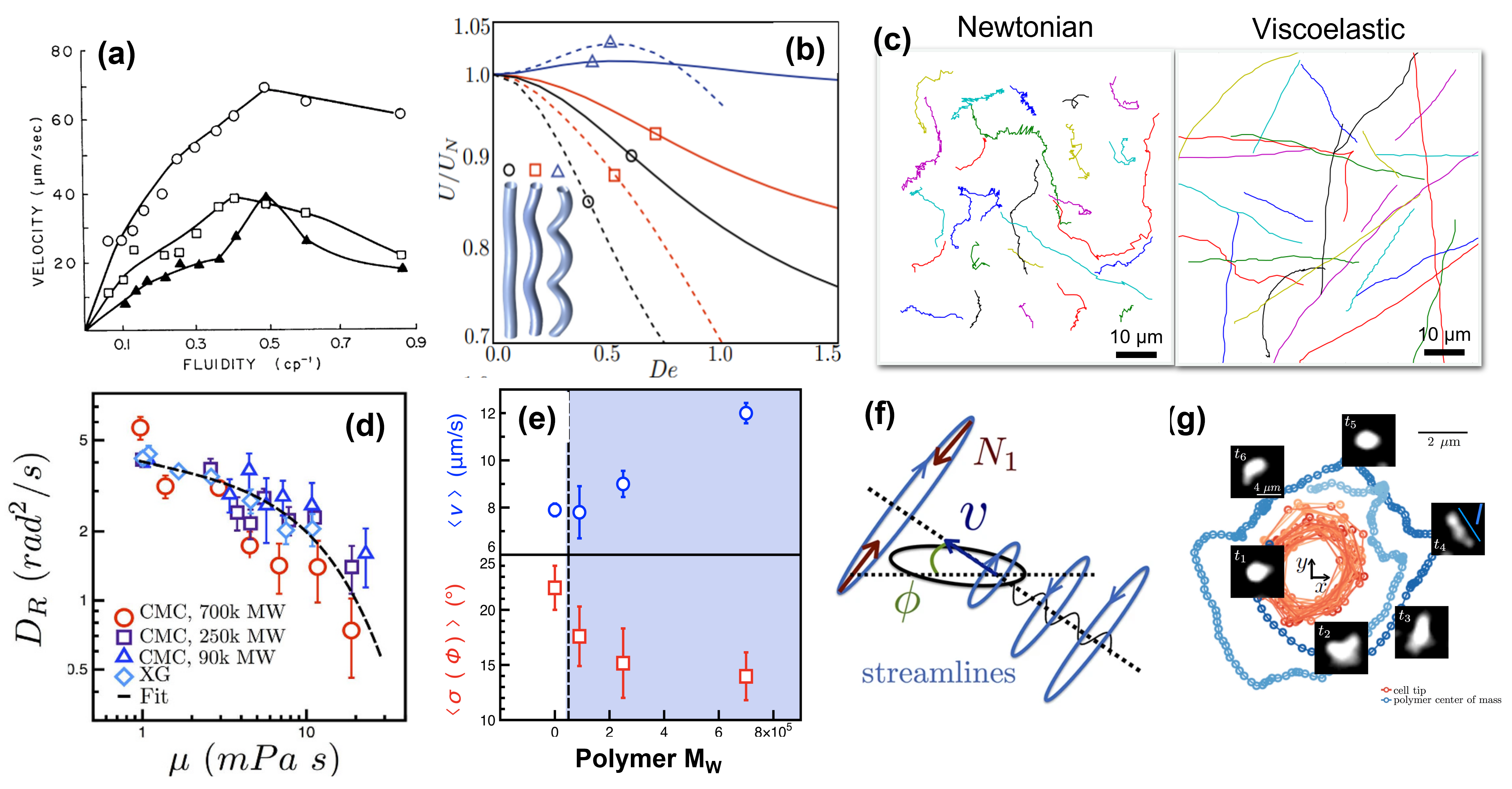}
 \caption{ \textit{E. coli} in polymeric solutions. (a) Experiments by Schneider \& Doetsch show that cell swimming speed increases as fluid viscosity increases \cite{schneide.wr_effect_1974}. (b) Numerical simulations with helical bodies in VE fluids showing the effects of body geometry on propulsion \cite{Saverio_PRL_2013}. (c) Experimental cell trajectories showing the suppression of tumbling in VE fluids \cite{patteson2015running}. (d) Rotational diffusivity for different polymeric solutions indicate the tumbling suppression is a viscous effect. (e) Increase in cell speed and the concomitant decrease in cell wobbling as a function of elasticity for a fixed viscosity ($\mu=10$~cP). (f) Schematic of hoop stresses acting on \textit{E. coli} cell. (g) Snapshots of DNA molecules being stretched by \textit{E. coli}'s swimming action.}
 \label{Fig:ecoli}
\end{figure*}

Nearly 50 years ago, Schneider \& Doetsch investigated the effects of non-Newtonian fluid viscosity on the swimming behavior of \textit{E. coli} \cite{schneide.wr_effect_1974}. Surprisingly, they found an increased in cell swimming speed with increasing fluid viscosity in aqueous solutions of poly-vinyl-pyrrolidone (PNP, $M_n = 360$ kDa) and of methyl-cellulose (MC, $M_n$ unknown); $M_n$ is the number-averaged molecular weight. The data reproduced from their original manuscript in Fig.~\ref{Fig:ecoli}(a) shows an increase in bacterium swimming speed with polymer concentration and a peak; note that the abscissa is in unis of inverse viscosity. It was argued at the time that \textit{E. coli} was able to move faster in polymeric solutions because they swim through polymer network pores, and thus only experience the solvent viscosity \cite{berg_movement_1979}. Numerical simulations based on this argument was able to reproduce some of the experimental results \cite{magariyama2002mathematical}. As discussed in \cite{Martinez_Poon_Ecoli_PNAS}, the proposed mechanism is not physical; the estimated pore size for the polymers used in the experiments are far too small, approximately 80 nm for PNP ($M_w=10^{6}$ kDa) assuming random coil, for an \textit{E coli} cell of cross section $\approx 1~\mu$m to move through it, among other issues. That prompt Martinez, Morozov, Poon, and colleagues \cite{Martinez_Poon_Ecoli_PNAS} to revisit the experiments of Schneider \& Doetsch. They carefully prepared a fresh set of fluids using the same type of polymers as the original study. They found that, for low $M_w$ polymers, \textit{E coli} swimming kinematics can be explained by Newtonian hydrodynamics alone. The authors argued that impurities in the polymeric solutions may have been responsible for the increased in cell swimming speed with $\mu$ observed by Schneider \& Doetsch. They showed that only the case with the highest $M_w$ polymeric solution (near overlap concentration) showed an increase in swimming speed due to local shear-thinning effects \cite{Martinez_Poon_Ecoli_PNAS}. They developed a minimal model that captures their experiments in the Newtonian (dilute) and shear-thinning (semi-dilute) regimes remarkably well. 

The effects of fluid elasticity on \textit{E. coli} swimming behavior, however, were less clear. Simulations of helical structures in viscoelastic fluids (Oldroyd-B model) show that elastic stresses can either enhance or hinder the structure propulsion speed and efficiency, depending on geometry (pitch, radius) and rotation rate \cite{Saverio_PRL_2013, Li_Spagnolie_PhysFluids_2015} (Fig.~\ref{Fig:ecoli}b). A local maximum in propulsion speed as a function of $De$ is found in the simulations, a result that is similar to experiments in scaled-up mechanical systems \cite{liu_force-free_2011}. The question is whether these findings translate to living microorganism.   

To address the question above, our laboratory performed a systematic experimental investigation on the effects of fluid elasticity on the swimming behavior of \textit{E. coli} \cite{patteson2015running}. Similar to the studies with \textit{C. reinhardtii}, experiments are performed in a thin film using Newtonian and polymeric solutions. Polymeric solutions are prepared using high molecular weight polymers ($M_w = 10 \times 10^6$) and are dilute, $c<c^*$ where $c^*$ is the overlap concentration; for more details on methods and protocols, please see \cite{patteson2015running}. Figure~\ref{Fig:ecoli}(c) shows \textit{E. coli} trajectories in both Newtonian and VE fluids using tracking techniques. While the typical run-and-tumble dynamics is observed in the Newtonian case, we find a very different behavior in VE fluids -- tumbling is suppressed as cell trajectories become more ballistic. It turns out, however, that this tumbling suppression is not due to fluid elasticity. Figure~\ref{Fig:ecoli}(d) shows \textit{E. coli} rotational diffusivity as a function of fluid viscosity $\mu$ for several polymeric solutions of different $M_w$ (and thus elasticity levels). The data show minimal differences among the different fluids suggesting that the suppression of tumbling (or rotational diffusivity) is a viscous effect.  

Fluid elasticity does seem to affect the \textit{E. coli} swimming speed. Figure~\ref{Fig:ecoli}(e) shows cell swimming speed (top) as a function of polymer molecular weight ($M_w$) for a fixed viscosity $\mu=10$ cP. The data shows a clear increase in cell as polymer $M_w$ (or elasticity) is increased, even though the fluid viscosity is kept relatively constant by adjusting polymer concentration below $c^*$. This means that the increase in speed is not a viscous effect and likely due to elasticity. Intriguingly, the enhancement in cell swimming speed is accompanied by a decrease in cell wobbling, as shown in Fig.~\ref{Fig:ecoli}(e, bottom). Cell wobbling are oscillations of the cell body along its path, a behavior typical of swimming \textit{E. coli}. These two-dimensional lateral oscillations of the cell body are projections of the cell’s three-dimensional helical trajectory. The bacterium \textit{E. coli} can wobble by as much as 20 to 30 degrees about its path centerline. The question is whether the decrease in wobbling is mechanistically related to the increase in bacteria speed in VE fluids. One could argue that the presence of curved streamlines in the \textit{E. coli} velocity field (Fig.~\ref{Fig:ecoli}f) could lead to polymer stretching and the production of elastic hoop stresses in a mechanism that is similar to the one responsible for the rod-climbing effect. As discussed in the introduction, these stresses points inward in the radial direction ($r$) towards the cell body and perpendicular to the cell’s swimming direction. These hoop stresses then cause the cell body to align with the projected direction of motion, thus reducing wobbling. Visualization of individual fluorescently labeled DNA polymers reveals that the flow generated by individual \textit{E. coli} is sufficiently strong to stretch polymer molecules and induce local elastic stresses in the fluid (Fig.~\ref{Fig:ecoli}g). Hence, we believe that hoop stresses are responsible for suppressing cell wobbling, which in turn leads to faster cell swimming speeds. 

In summary, these results show how local shear-thinning effects \cite{Martinez_Poon_Ecoli_PNAS} and elastic stresses \cite{patteson2015running} can significantly affect the swimming behavior of \textit{E. coli}. Despite progress, the mechanism responsible for suppression of tumbling and the changes in cell speed in VE fluids are still being debated; there is no consensus just yet. For example, recent numerical simulations find that elasticity can indeed lead to an increase in cell swimming speed due to azimuthal swirl in its gait that decreases the extensional wake behind the swimmer \cite{binagia_phoa_housiadas_shaqfeh_2020}, while an experimental investigation show that wobbling is indeed reduced by normal stresses but its not the main cause for increase in swimming speed. Rather, speed enhancement is due to shear-thinning effects similar to \cite{Martinez_Poon_Ecoli_PNAS}. Very recently an intriguing study by X. Cheng and colleagues showed that \textit{E. coli} can swim faster in suspensions of colloidal particles \cite{Cheng_Colloidal_Nature2022}. They argued (and demonstrated) that as bacteria cells swim near particles, they experience a torque that aligns the flagella with their body leading to faster swimming. A similar mechanism may be at play when cell move near polymer molecules. This study, as well a recent numerical simulation \cite{Yeomans_NatPhys_2019}, shows how nonlocal effects must be consider particularly as as the length scale of fluid microstructure is of the same order as the cell length scale; the continuum approach breaks down under those conditions. clearly, there is still much that we do not know regarding the effects of fluid rheology on the swimming behavior of \textit{E. coli}.  

\section{Conclusions \& Outlook}
Swimming in complex fluids is a rich, nonlinear problem that still is not fully understood. The two-way coupling between swimmer kinematics and fluid rheological properties can give rise to many unexpected behaviors, as shown here. In some instances, fluid rheology can aid propulsion but in others it may be detrimental. It is, therefore, difficult to make general statements regarding propulsion speed and/or energy expenditure because much depends on how the swimmer interacts with the polymers and particles in the fluid. Nevertheless, the field has made much progress in characterizing/modeling such interactions with the goal of developing general understanding of motility. Opportunities for those interested in joining the community are still plenty. I will discuss a few below. 

Perhaps one of the most outstanding questions is whether the gait or kinematic changes observed in the experiments is a passive or active response (or a combination of both). For example, we show that elastic stresses can significantly affect and change the beating waveform of \textit{C. reinhardtii}, relative to what is commonly observed in simple, Newtonian fluids (Fig.~\ref{Fig:Chlamy}). Yet, we are not sure whether the alga cell is actively responding to fluid stresses. It has been shown in experiments \cite{sznitman_effects_2010}, simulations \cite{qin_chuanbin_2017}, and analysis \cite{lauga_eloy_2013} that organisms' motility behavior can vary widely depending on flagellum/organism (passive) material properties. One avenue to address this question (active vs passive response) is to combine experimental data with numerical simulations \cite{qin_chuanbin_2017} in order to decouple swimming kinematics from fluid rheological effects. Moving forward, it is desirable to include accurate models for the flagellum active forces (e.g., dynein motor activity), for instance, in the fluid-swimmer formulation. Could these types of formulations capture the emerging flagellum waveforms? 

Albeit described only briefly \cite{Martinez_Poon_Ecoli_PNAS, Yeomans_NatPhys_2019, Cheng_Colloidal_Nature2022}, the importance of resolving nonlocal effects cannot be understated. While there is enough separation of scales for microorganism swimming in Newtonian fluids, that may not be the case for fluids containing polymers and/or solids. This can be quantified by the Knudsen number, $K_n=L_f/L_s$, where $L_f$ and $L_s$ are the characteristic length scales of the fluid and swimmer, respectively. Consider for example \textit{E. coli} $L_s \sim \mathcal{O}(1)\mu$m moving in water $L_f \sim \mathcal{O}(0.1)$nm. In such case $K_n ~ \mathcal{O} (10^{-4})$, and the system can be adequately described by the continuum approach. The picture is different even in polymeric solution; polymer radius of gyration,$R_g$, for high $M_w$ molecules can be as high as 300 nm. Then, $Kn=0.3$ for the same \textit{E. coli} indicating that a molecular, statistical approach may be more adequate to describe such swimmer \cite{Arezoo_JNNFM, Yeomans_NatPhys_2019}. Of course, a natural length scale to consider is the one associated with the velocity decay, $r_v$. Considering again the \textit{E. coli}, we can estimate $r_v$ by first noting that the velocity decays as $1/r^2$ (for a pusher dipole). The flagellum helix diameter, $a$, is approximately 0.25 $\mu$m and it rotates at an angular speed $\Omega$ of about 170 rad/s. We can estimate the velocity decay to 10$\%$ of the maximum speed next to the rotating flagellum ($a\Omega)$ to be $(a^2/r^2)a\Omega=0.1 a\Omega$. This gives a length scale $r_v = a\sqrt{10}=0.75~\mu$m. Hence, the $K_n=R_g/r_v = 0.4$. This is a similar result as using the $L_s \sim \mathcal{O}(1)\mu$m suggesting that the use of statistical approach is warranted. For a more comprehensive discussion on this topic, please see \cite{Saverio_Underhill_ARFM}. 

Finally, while there has been much progresses in understanding swimming of \textit{single} organisms in complex fluids, much less is now about their collective motion. Only a few investigation are available: numerical simulations predict that elasticity can significantly affect the size of clusters in non-dilute swimmer suspensions \cite{bozorgi_effect_2011, Li_PRL_2016}, while experiments with sperm show that polymers can even promote collective swimming \cite{tung_fluid_2017}. Recently, large oscillatory vortices were found in bacterial suspensions inside droplets containing viscoelastic fluids (DNA suspensions) \cite{Liu_Nature_2021}. It is still unclear, however, how polymers mediate microorganism hydrodynamic interactions and affect collective motion. 

The above are just a few areas in need of development. Needless to say that they come with a bit of bias from the author. I am hoping that after reading this article, the readers will have their own ideas on how to move our field forward.

\begin{acknowledgments}
I would be remiss if I did not acknowledge all current and past members of my research laboratory for their contributions to the field. In particular I would like to thank J. Sznitman, N. Keim, G. Juarez, M. Garcia, X. Shen, A. Patteson, A. Gopinath, B. Qin, and D. Gagnon for their intellectual insights and sweat. Finally, many thanks to B. Torres Maldonado and R. Ran for their help in putting this manuscript together. PEA is currently supported by NSF-DMR-1709763. 
\end{acknowledgments}

\bibliography{references_PRF.bib}

\end{document}